\documentclass[twocolumn, aps, superscriptaddress, prb, showpacs,longbibliography]{revtex4-1}
\usepackage{dcolumn}
\usepackage{amsmath,amssymb}
\usepackage{bm}
\usepackage{color}
\usepackage{url}
\usepackage{here}
\usepackage{braket}
\usepackage{multirow}

\usepackage{ulem}
\usepackage{color}
\DeclareRobustCommand{\delete}{\bgroup\markoverwith{\textcolor{red}{\rule[.5ex]{2pt}{0.4pt}}}\ULon}

\usepackage[pdftex]{graphicx}

\usepackage[pdftex]{pict2e,hyperref}

\begin{document}


\title {Fractional hinge and corner charges in various crystal shapes with cubic symmetry}
\author {Katsuaki Naito}
\affiliation{
Department of Physics, Tokyo Institute of Technology, 2-12-1 Ookayama, Meguro-ku, Tokyo 152-8551, Japan\\
}
\author {Ryo Takahashi}
\affiliation{
Department of Physics, Tokyo Institute of Technology, 2-12-1 Ookayama, Meguro-ku, Tokyo 152-8551, Japan\\
}
\author {Haruki Watanabe}
\affiliation{
Department of Applied Physics, University of Tokyo, Tokyo 113-8656, Japan\\
}
\author {Shuichi Murakami}
\affiliation{
Department of Physics, Tokyo Institute of Technology, 2-12-1 Ookayama, Meguro-ku, Tokyo 152-8551, Japan\\
}
\affiliation{
TIES, Tokyo Institute of Technology, 2-12-1 Ookayama, Meguro-ku, Tokyo 152-8551, Japan\\
}

\date{\today}

\begin{abstract}
Higher-order topological insulators host gapless states on hinges or corners of three-dimensional crystals.
Recent studies suggested that even topologically trivial insulators may exhibit fractionally quantized charges localized at hinges or corners. Although  most of the previous studies focused on two-dimensional systems,  in this work,  we take the initial step toward the systematic understanding of hinge and corner charges in three-dimensional insulators.
We consider five crystal shapes of vertex-transitive polyhedra with the cubic symmetry such as a cube, an octahedron and a cuboctahedron. 
We derive real-space formulas for the hinge and corner charges in terms of the electric charges associated with bulk Wyckoff positions.
We find that both the hinge and corner charges can be predicted from the bulk perspective only modulo certain fractions depending on the crystal shape, because the relaxation near boundaries of the crystal may affect the fractional parts. 
In particular, we show that a fractionally quantized charge $1/24$ mod $1/12$ in the unit of elementary charge can appear in a crystal with a shape of a truncated cube or a truncated octahedron. We also investigate momentum-space formulas for the hinge and corner charges. It turns out that the irreducible representations of filled bands at high-symmetry momenta are not sufficient to determine the corner charge. We introduce an additional Wilson-loop invariant to resolve this issue.
\end{abstract}

\maketitle

\section{Introduction}
Topological insulators are characterized by a bulk band gap and topological invariants formulated in terms of the Bloch wave functions\cite{RevModPhys.82.3045,RevModPhys.83.1057}. The bulk-boundary correspondence\cite{PhysRevB.74.195312,PhysRevB.76.045302,PhysRevB.78.195125,Ryu2010} then implies the presence of gapless excitations localized at the boundaries of topological insulators.
Three-dimensional topological insulators usually feature gapless modes on their two-dimensional surfaces
\cite{
PhysRevLett.95.146802,
PhysRevLett.98.106803,
bernevig2006quantum,
konig2007quantum}.
When all the surfaces are gapped but hinges or corners are gapless, the insulator is said to possess a higher-order topology
\cite{
PhysRevLett.106.106802,
HOTIbismuth,
HOTI2018,
PhysRevB.98.081110,
PhysRevResearch.2.012067,
PhysRevLett.124.036803,
PhysRevLett.119.246402,
watanabe2020fractional,
PhysRevResearch.2.043131,
PhysRevB.103.205123,
PhysRevB.103.165109,
PhysRevResearch.1.033074,
PhysRevB.99.245151,
PhysRevB.102.165120,
PhysRevB.48.4442,
PhysRevB.101.115115,
PhysRevB.96.245115}.

In contrast, electronic excitation spectrum of topologically trivial insulators is completely gapped including hinges and corners.
Electrons in these insulators occupy exponentially localized Wannier orbitals, which usually resemble atomic orbitals.  Thus topologically trivial insulators are also called atomic insulators (AIs).  Recent studies found that the boundaries of AIs are not completely featureless; some AIs feature fractionally quantized charges on their corners. The most of previous studies of fractional corner charges have been limited to two-dimensional systems, except for Refs.~[\onlinecite{watanabe2020fractional},\onlinecite{PhysRevB.102.165120}] in which the fractional corner charge of a cubic crystal has been investigated.

The fractional corner charges of two-dimensional systems have been understood in terms of filling anomaly\cite{PhysRevB.99.245151}.
In the presence of a point-group symmetry such as inversion symmetry and $n$-fold rotation symmetry, the possible positions of ions and electronic Wannier orbitals obey some symmetry constraints. If some Wannier orbitals of filled states of AIs are not located at ionic positions, the AI is classified as an obstructed atomic insulator (OAI), which is characterized by charge imbalance associated with each Wyckoff position\cite{bradlyn2017topological,PhysRevB.97.035139,po2017symmetry,cano2021band}.  
When the point-group symmetry is strictly required including its boundary, a finite crystal of an OAI under an open boundary condition cannot be electrically charge neutral because of the mismatch between the total number of electrons and ions in the system. This charge imbalance is called filling anomaly. The fractional charge localized at a corner can be deduced by dividing the filling anomaly by the number of corners, which are related by symmetry
\cite{
PhysRevB.96.245115,
watanabe2020fractional,
PhysRevResearch.2.043131,
PhysRevB.103.205123,
PhysRevB.103.165109,
PhysRevResearch.1.033074,
PhysRevB.99.245151,
PhysRevB.102.165120,
PhysRevB.48.4442}.

In two dimensions, formulas for the corner charge are expressed in terms of irreducible representations of the little group at high-symmetry momenta\cite{PhysRevResearch.1.033074,
PhysRevB.103.205123}.
These formulas predict the fractional parts of the corner charges from the bulk band structure without referring to the details of the surface termination. Therefore they can be interpreted as the bulk-corner correspondence of topologically trivial insulators. In other words, the fractional parts of the corner charges in two-dimensional insulators are insensitive to possible relaxations of electronic states and ions near the boundary, which cannot be inferred from the bulk band structure alone.

There are two main sources of complications in extending the above results to three-dimensional insulators. One complication comes from the existence of various distinct crystal shapes for the same point group symmetry in three dimensions. This is different from two-dimensional cases, where a regular $n$-polygon is basically the unique shape which preserves the $n$-fold rotation symmetry ($n=3,4,6$) and has straight edges.  In this work, we focus on the systems preserving the point group symmetry $O$, corresponding to the space group $P432$ (No.~207). We discuss five different crystal shapes for this symmetry: a cube, an octahedron, a truncated cube, a cuboctahedron and a truncated octahedron (see Figs.~\ref{unitcell} (a)-(e)). These shapes are vertex-transitive polyhedra, in which all the corners are related by the point group symmetry. This property is a necessary condition for the quantization of the corner charge.  

The other complication comes from the charge neutrality conditions at the boundary. To define the corner charge unambiguously for three-dimensional crystals, not only the bulk and the surfaces but also the hinges must be charge neutral. In the studies so far
\cite{PhysRevResearch.2.043131,
PhysRevB.102.165120}, the hinge charges have been calculated as the corner charge of two-dimensional layer which constitutes the three-dimensional crystal.  In this paper, we derive formulas for the hinge charge of the three-dimensional systems, including those which cannot be formed by stacking of two-dimensional layers such as an octahedron.

In this work, we first derive formulas for the hinge and corner charges in terms of the charge imbalance at each Wyckoff position in the bulk. We discuss five crystal shapes mentioned above.
Our strategy is to determine the filling anomaly of the system by counting the total numbers of electrons and ions given the positions of ions and Wannier centers of the electrons. From the formula of the filling anomaly, we can extract formulas of the hinge charge and corner charge. Next, we identify ambiguities of the hinge and corner charges originating from the relaxation of electronic states and ionic positions near the boundary. We also explore momentum-space formulas for the hinge and corner charges. We obtain a formula for the hinge charge written in terms of the irreducible representations (irreps) at high-symmetry momenta in the Brillouin zone (BZ) based on the elementary band representation (EBR) matrix method developed in Refs.~[\onlinecite{PhysRevB.103.165109},\onlinecite{cano2021band}]. However, we find that this approach fails to determine the corner charge because the information on the irreps at high-symmetry momenta is not sufficient to fix the electronic Wannier centers of occupied bands. We resolve this problem by introducing an additional Wilson-loop invariant.

This paper is organized as follows. In Sec.~\ref{Sec2}, we summarize charge neutrality conditions for the bulk and surfaces with cubic symmetry in terms of the occupation numbers of each Wyckoff position in the bulk. In Sec.~\ref{Sec3}, we derive real-space formulas for the hinge and corner charges. In Sec.~\ref{formulation by topological invariants}, we reformulate the results obtained in the previous sections in terms of the EBR matrix and a Wilson-loop invariant. Conclusion is given in Sec.~\ref{CaD}.

\section{charge neutrality conditions for the bulk and surfaces in terms of the bulk Wyckoff positions}
\label{Sec2}
\begin{figure}
\centerline{\includegraphics[width=9cm,clip]{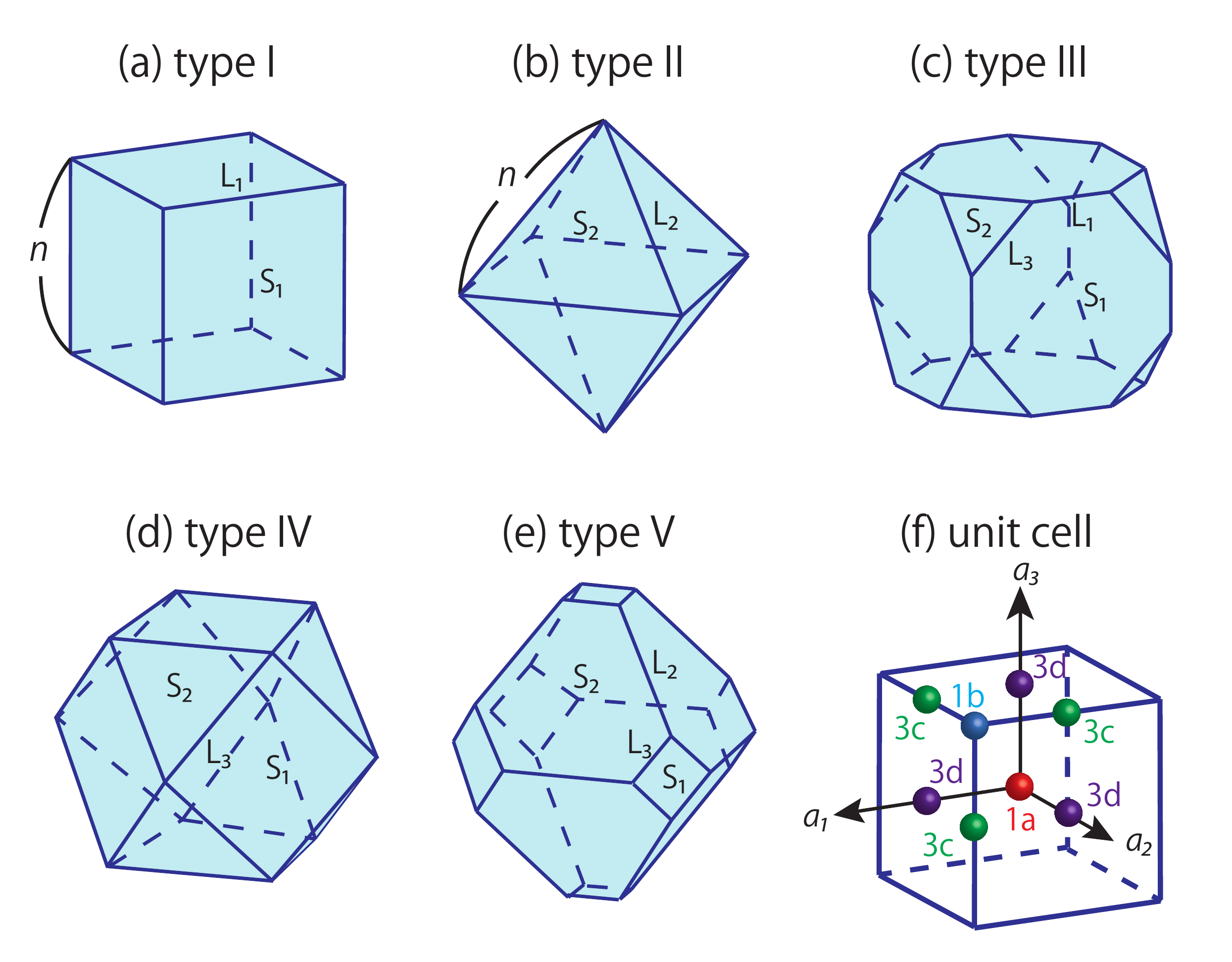}}
\caption{(a)-(e) Five crystal shapes considered in this work, all of which are vertex-transitive polyhedra. Their center is placed at Wyckoff position $1a$. (a) type I: a cube. (b) type I\hspace{-1pt}I: a regular octahedron. (c) type I\hspace{-1pt}I\hspace{-1pt}I: a truncated cube. (d) type I\hspace{-1pt}V: a cuboctahedron. (e) type V: a truncated octahedron. The surfaces $S_{1}$ and $S_{2}$ have Miller index $\{100\}$ and $\{111\}$, respectively. The hinge $L_{1}$ is an intersection of two $\{100\}$ surfaces. The hinge $L_{2}$ is an intersection of two $\{111\}$ surfaces. The hinge $L_{3}$ is an intersection of a $\{100\}$ surface and a $\{111\}$ surface. (f) Wyckoff positions in the space group $P432$. The points with the same colors belong to the same
Wyckoff positions. $1a$, $1b$, one of $3c$ and one of $3d$ are positioned at $(0,0,0)$, $\frac{\bm{a}_{1}+\bm{a}_{2}+\bm{a}_{3}}{2}$, $\frac{\bm{a}_{1}+\bm{a}_{2}}{2}$ and $\frac{\bm{a}_{1}}{2}$, respectively. The blue cube represents the unit cell.}
\label{unitcell}
\end{figure}

For our final goal of calculating the hinge charge and the corner charge in cubic systems in terms of the charge imbalance at each Wyckoff position in the bulk, in this section, we derive the charge neutrality conditions for the bulk and surfaces for finite-sized crystals. 
There are 32 point groups in three dimensions and we consider the cubic point group symmetry $O$, which is the point group for the space group $P432$. In particular we consider five crystal shapes, types I-V illustrated in Figs.~\ref{unitcell} (a)--(e).
We assume that the systems we consider are topologically trivial in the sense that they are adiabatically connected to an atomic limit. 
We also assume that the excitation energy spectrum of the system is completely gapped including the boundaries.

To calculate the corner charge, we note that electric charges in insulators can be assigned to either ions or Wannier orbitals of occupied electronic bands. An ion is made of a nucleus and core electrons, and the ionic charge is given by the sum of their electric charges, which, by definition, is integral. Note that the ionic charge defined in this way is different from the effective charge of the ions in ionic crystals; such an effective charge is usually non-integral. Moreover, in topologically trivial insulators, Wannier orbitals are exponentially localized, and the integral charge of electrons in the Wannier orbital can be assigned to the Wannier center.

Let $w(=a,b,c,d)$ be one of the maximal Wyckoff positions in the cubic unit cell of the cubic lattice as shown in Fig.~\ref{unitcell} (f). 
Let $n_{w}$ denote the number of Wannier functions centered at a Wyckoff position $w$ in the bulk, and $m_{w}$ denote the total charge of ions measured in the unit of elementary charge $e(>0)$ at a Wyckoff position $w$ in the bulk.  We then define $\Delta w$ to be the difference between them: 
\begin{equation}
\Delta w=n_{w} - m_{w}.
\end{equation}
By definition, $\Delta w$ is always an integer.
Here, the space group $P432$ allows Wyckoff positions $1a$, $1b$, $3c$, $3d$, $6e$, $6f$, $8g$, $12h$, $12i$, $12j$ and $24k$, and we can restrict ourselves to the maximal ones, $1a$, $1b$, $3c$ and $3d$, because the others can be reduced to the maximal ones via continuous transformations. Throughout this work, we assume that the center of a crystal is at Wyckoff position $1a$ unless otherwise stated.

In order to derive a corner charge formula, we review the definition of filling anomaly\cite{PhysRevB.99.245151}. In some bulk insulators, we cannot make the system charge neutral as long as the system preserves the required symmetry. In such cases, we have to add or remove electrons from charge neutrality to make the system insulating including the boundaries. This number of extra electrons is called  filling anomaly. Since the total charge of ions measured in the unit of elementary charge is equal to the number of electrons under charge neutrality, we can express filling anomaly $\eta_{n}$ of the finite-sized crystal as 
\begin{equation}
\label{FA}
\eta_{n}=N_{n}^{\text{electron}}-N_{n}^{\text{ion}},
\end{equation}
where $N_{n}^{\text{electron}}$ is the total number of electrons in the finite-sized crystal preserving $O(432)$ symmetry, and $N_{n}^{\text{ion}}$ is the total charge of all the ions measured in the unit of elementary charge in the same setup. Here, the parameter $n$ characterizes the system size (e.g., the number of unit cells along one hinge of the crystal).

To proceed, let us focus on type I and type I\hspace{-1pt}I crystals, in which all faces and hinges are equivalent (i.e., related to each other by the point group symmetry $O$). We tentatively assume that there is no surface reconstruction so that the periodicities of the hinge and the surface reflect that of the bulk. This implies that the crystal shape should have straight hinges and flat surfaces. Under this assumption, the filling anomaly for a finite-sized crystal with $n$ hinge periods [see Fig.~\ref{unitcell} (a) and (b)] can be expanded in a power of $n$: 
\begin{equation}
\label{FAexpanded}
\eta_{n}=\alpha_{3}n^{3}+\alpha_{2}n^{2}+\alpha_{1}n+\alpha_{0}.
\end{equation} 
The  terms on the right hand side can be interpreted as contributions from the bulk, surfaces, hinges, and corners. The coefficients $\alpha_{3}$ and $\alpha_{2}$ are, respectively,  related to the bulk charge density $\rho_{\text{bulk}}$ (per bulk unit cell) and the surface charge density $\sigma_{\text{sur}}$ (per surface unit cell) via $\alpha_{3}=\alpha'_{3}\rho_{\text{bulk}}/(-e)$ and $\alpha_{2}=\alpha'_{2}\sigma_{\text{sur}}/(-e)$ where $\alpha'_{3}=\alpha'_{2}=1$ for type I and $\alpha'_{3}=\frac{4}{3}$ and $\alpha'_2=4$ for type I\hspace{-1pt}I. 
When both the bulk and surfaces are charge neutral, we can proceed to the hinge and corner charges encoded in $\alpha_{1}$ and $\alpha_{0}$, as we discuss in detail in the next section.
In the remainder of this section, we derive the charge neutrality conditions for the bulk and surfaces.

First of all, the bulk charge density is identified as $(\Delta a+\Delta b+3\Delta c+3\Delta d)\times (-e)$ by counting charges on the Wyckoff positions included in the bulk unit cell as shown in Fig.~\ref{unitcell} (f). 
Thus, the charge neutrality condition for the bulk is given by
\begin{equation}
\label{BCNC}
\rho_{\text{bulk}}=-e(\Delta a+\Delta b+3\Delta c+3\Delta d)=0.
\end{equation}
This is equivalent to assuming $\alpha_{3}=0$ in Eq.~(\ref{FAexpanded}).

Next let us investigate the surface charge. According to the modern theory of polarization \cite{PhysRevB.47.1651, PhysRevB.48.4442}, the bulk polarization is given by
\begin{equation}
\label{BP}
\bm{P}_{\text{bulk}}=\frac{-e}{2a^3}(\Delta b+\Delta d)(\bm{a}_{1}+\bm{a}_{2}+\bm{a}_{3})\ (\text{mod}\ \frac{e}{a^{3}}\bm{R}).
\end{equation}
Here, $\bm{a}_{1}, \bm{a}_{2}$ and $\bm{a}_{3}$ are primitive lattice vectors in the cubic unit cells: $\bm{a}_{1}=a\hat{\bm{x}}$, $\bm{a}_{2}=a\hat{\bm{y}}$ and $\bm{a}_{3}=a\hat{\bm{z}}$, where $a$ is the lattice constant. $\bm{R}=\sum_{i=x,y,z}^{}m_{i}\bm{a}_{i} \, (m_{i}\in \mathbb{Z})$ is a lattice vector. 
The surface charge density $\sigma_{\text{sur}}$ with its normal vector $\bm{n}$ is given in terms of the bulk polarization as 
\begin{equation}
\label{SCD}
\sigma_{\text{sur}}=\bm{P}_{\text{bulk}}\cdot \bm{n}\, s_{\text{sur}}\ (\text{mod}\ e),
\end{equation}
where $s_{\text{sur}}$ is the area of the surface unit cell. 
For example, the type I crystal shown in Fig.~{\ref{unitcell} (a)} has $\{100\}$ surfaces. The surface charge density is given by substituting $\bm{n}=(1,0,0)$ and $s_{\text{sur}}=a^2$ to Eq.~(\ref{SCD}): 
\begin{equation}
\label{surfacechargedensity}
\sigma_{\text{sur}}=-\frac{\Delta b +\Delta d}{2}e\ (\text{mod}\ e).
\end{equation}
Thus, the charge neutrality condition for the surface is
\begin{equation}
\label{SCNC2}
\Delta b +\Delta d\ \equiv 0\ (\text{mod}\ 2).
\end{equation}
On the other hand, the type I\hspace{-1pt}I crystal in Fig.~\ref{unitcell} (b) has $\{111\}$ surfaces. 
The surface charge density, obtained by substituting $\bm{n}=(1,1,1)/\sqrt{3}$ and $s_{\text{sur}}=\sqrt{3}a^2$ to Eq.~(\ref{SCD}), turns out to be the same as Eq.~(\ref{surfacechargedensity}). 
Actually, the charge neutrality condition for surfaces is the same in all the five crystal shapes considered in this paper as shown later. 

Note that even when Eq.~(\ref{SCNC2}) is satisfied, the surface charge density can still be nonzero and may be an integer multiple of $e$. If it is nonzero, we always introduce charges to the surface that precisely cancel the surface charge density. In this way, we assume $\alpha_{2}=0$ in Eq.~(\ref{FAexpanded}), guaranteeing that the hinge charge is well-defined.

\section{hinge charge and corner charge formulas in terms of the bulk Wyckoff positions}
\label{Sec3}
In this section, we derive formulas for the hinge charge density and the corner charge in 3D cubic systems in terms of the bulk Wyckoff positions for five crystal shapes illustrated in Figs.~\ref{unitcell} (a)-(e). 

Before discussing each type of a crystal, we outline how to calculate the hinge and corner charges in types I and I\hspace{-1pt}I.
Henceforth, we assume the charge neutrality in the bulk [Eq.~(\ref{BCNC})] and on the surfaces [Eq.~(\ref{SCNC2})].
These assumptions make the first and second terms on the right hand side in Eq.~(\ref{FAexpanded}) vanish:
\begin{equation}
\label{FA2}
\eta_{n}=\alpha_{1}n+\alpha_{0}.
\end{equation} 
This expression can be interpreted as a sum of the total hinge charge $\alpha_{1}n$ and the total corner charge $\alpha_{0}$, because the extra charge exists only on hinges with $n$ periodicities and corners under the assumptions. Let $\lambda_{\text{hinge}}$ be the hinge charge density (per hinge unit cell). Since there are equivalent twelve hinges both in type I and type I\hspace{-1pt}I, the total hinge charge is $12\lambda_{\text{hinge}}n$. Thus, we get
\begin{equation} 
\label{hinge_1_2}
 \lambda_{\text{hinge}}=-\frac{\alpha_{1}}{12}e,
\end{equation}
for type I and type I\hspace{-1pt}I crystals. The hinge charge density in the other types of crystals can be calculated by using this result as discussed later.  

The corner charge is well-defined only when the hinges are charge neutral, in addition to the bulk and surfaces, i.e., $\alpha_{1}=0$. Given that all the corners are related by point group symmetry, the excess charge is eventually distributed equally on each corner under these conditions. Thus, the charge localized at a single corner $Q_{\text{corer}}$ is given by
\begin{equation}
\label{corner_general}
Q_{\text{corner}}=-\frac{\alpha_{0}}{N_{\text{corner}}}e,
\end{equation}
where $N_{\text{corner}}$ is the number of symmetry-related corners in the crystal, which depends on the crystal shape. The values of $N_{\text{corner}}$ for the crystal type I-V are listed in Table~\ref{real-space-formulas}.

\begin{figure}[h]
\centerline{\includegraphics[width=8cm,clip]{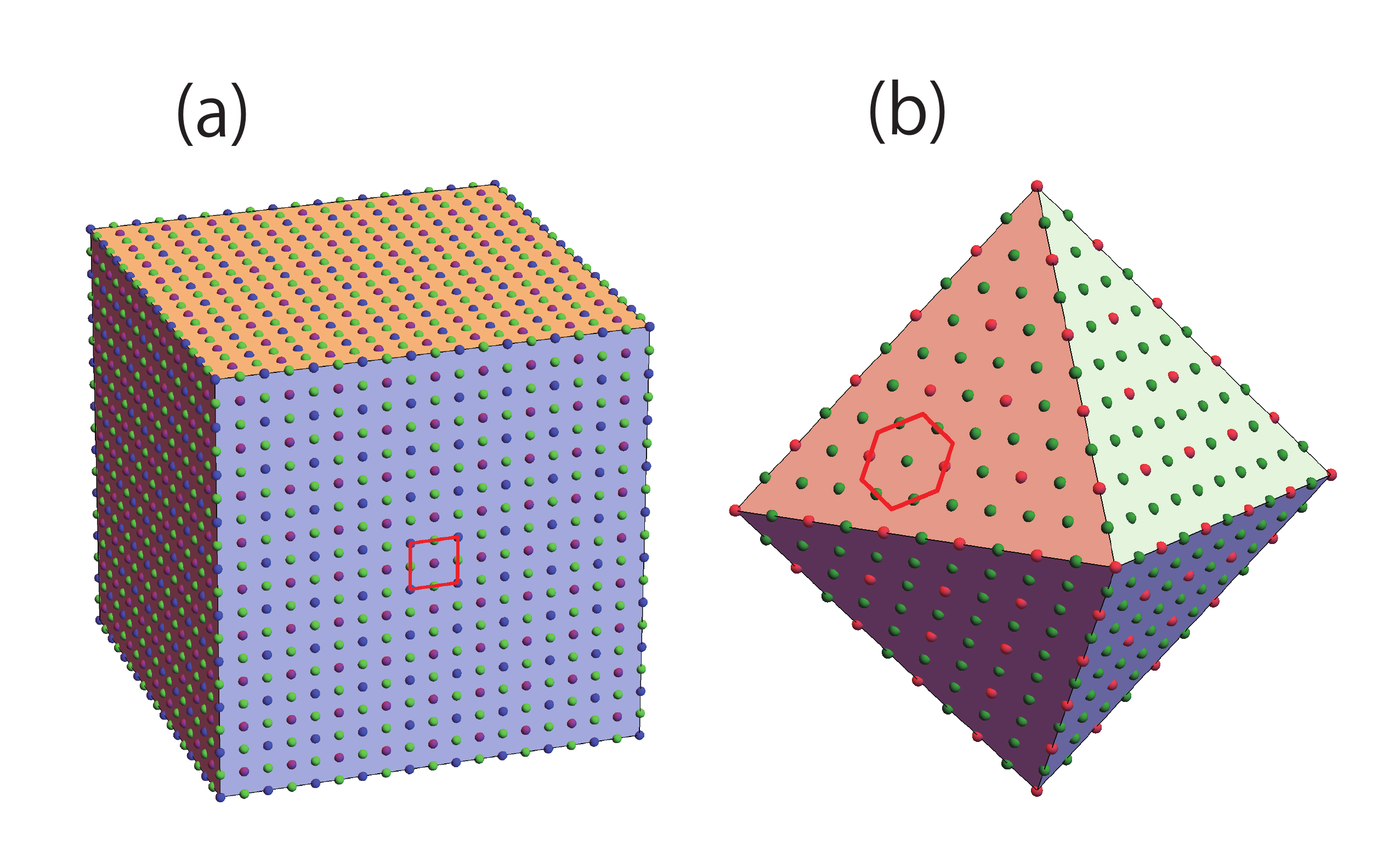}}
\caption{Perfect crystals for (a) the type I ($n=9$) and (b) the type I\hspace{-1pt}I ($n=5$) shapes. The red, blue, green and purple spheres represent Wyckoff positions $1a$, $1b$, $3c$ and $3d$, respectively. The areas enclosed by the red lines represent surface unit cells.}
\label{perfectfig1}
\end{figure}

What remains to be done is to calculate $\alpha_{0}$ and $\alpha_{1}$ for each type of a crystal. 
We begin with the simplest case of perfect crystals that comprise exactly identical unit cells with bulk electronic states and ionic positions even near the boundaries, as illustrated in Fig.~\ref{perfectfig1}. We calculate the filling anomaly for perfect crystals of type I and type \mbox{I\hspace{-1pt}I} and derive the formulas for the hinge charge density and the corner charge in Secs.~\ref{sectype1} and \ref{sectype2}.

Note that perfect crystals are only the special cases belonging to these types of crystals.
Generally, crystals in the same shape can have different electronic states and ionic positions near the boundaries, even when the crystals share the same bulk ones. Such difference cannot be fixed from the bulk perspective, but it modifies the values of $\alpha_{0}$, $\alpha_{1}$ and $\alpha_{2}$ in the filling anomaly formula in Eq.~(\ref{FAexpanded}). This fact should be understood as a limitation of predicting the surface, hinge, and corner charges from the bulk charge distribution. In the following, we refer to the part of the boundary charge that is affected by surface reconstruction as the ambiguity of the boundary charge. For example, the ambiguity of the surface charge density is its integer part (in the unit of $e$) as in Eq.~(\ref{SCD}) from the modern theory of polarization\cite{ PhysRevB.48.4442,PhysRevB.47.1651}. We derive the ambiguities of the hinge and corner charges in Sec.~\ref{ApA}. 
Making use of these results, we also derive the hinge charge density and the corner charge for type I\hspace{-1pt}I\hspace{-1pt}I, type I\hspace{-1pt}V and type V in Secs.~\ref{sectype3}, \ref{sectype4} and \ref{sectype5}, respectively.
We note that the choice of perfect crystals is not unique, and the hinge charge densities and corner charges depend on the choice of perfect crystals. This dependence is a part of the ambiguity of the boundary charge discussed above.

\subsection{type I: cube}\label{sectype1}
Here we discuss a crystal in the shape of a cube with $n$ unit cells along each hinge, as shown in Fig.~\ref{unitcell} (a). 
From direct calculation, we obtain the filling anomaly for the perfect crystal of type I as shown in Fig.~\ref{perfectfig1} (a): 
\begin{eqnarray}
\eta_{n}^{\text{perfect,\ type\ I}}&=&(\Delta a+\Delta b+3\Delta c+ 3\Delta d)n^{3}\nonumber \\ 
&&+3(\Delta b+\Delta d+2\Delta c)n^{2}\nonumber \\ 
&&+3(\Delta b+\Delta c)n+\Delta b.
\label{perfect_filling1}
\end{eqnarray}
The first and second terms in the right hand side can be dropped when the charge neutrality conditions for the bulk and surfaces [Eqs.~(\ref{BCNC}) and (\ref{SCNC2})]  are satisfied. Thus, from Eq.~(\ref{hinge_1_2}), we obtain the hinge charge density
\begin{eqnarray}
\label{HC1}
\lambda_{\text{hinge}}^{\text{type\ I}}&=&-\frac{\Delta b+\Delta c}{4}e \nonumber \\
&=&-\frac{\Delta a+\Delta d}{4}e\ (\text{mod}\ e).
\end{eqnarray}
Hence, the charge neutrality condition for the hinge is
\begin{equation}
\label{HCNC1}
\Delta a +\Delta d\ \equiv 0\ (\text{mod}\ 4).
\end{equation}
When Eq.~(\ref{HCNC1}) is satisfied in addition to Eqs.~(\ref{BCNC}) and (\ref{SCNC2}), we find the corner charge from Eq.~(\ref{corner_general})
\begin{equation}
\label{CC1}
Q_{\text{corner}}^{\text{type\ I}}=-\frac{\Delta b}{8}e=-\frac{\Delta a}{8}e\ \left(\text{mod}\ \frac{e}{4}\right).
\end{equation}
Here we used the relation $\Delta a\equiv \Delta b\equiv \Delta c\equiv \Delta d$ (mod 2) under Eqs.~(\ref{BCNC}), (\ref{SCNC2}) and (\ref{HCNC1}). 

Let us rationalize the ambiguities of $e$ and $e/4$ in Eqs.~(\ref{HCNC1}) and (\ref{CC1}).
They originate from possible relaxation of electronic states and ionic positions near the boundaries, which may be understood as decoration of boundaries with lower dimensional objects in a symmetric manner without affecting the bulk of the crystal. Here, for simplicity, we consider the ambiguities of the hinge and corner charges by a special approach where lower dimensional systems respecting the required symmetry are attached to the boundaries of the three-dimensional system. It turns out that the resulting ambiguities are the same with those obtained from general discussions in Sec.~\ref{ApA}. For example, each hinge in the cube is an intersection of two $C_{4}$-symmetric squares. We can attach a single layer of $C_{4}$-symmetric squares with a quantized polarization $\bm{P}\equiv (m/(2a),m/(2a))e\ (\text{mod}\ e/a)$ on every surface ($m$: an integer), which changes the hinge charge by $me$. Similarly, the corner charge is affected by attaching $C_{4}$-symmetric 2D systems on surfaces, $C_{2}$-symmetric 1D systems on hinges, and 0D systems at corners.
In particular, 2D systems with fractionally quantized corner charge $me/4$ change the corner charge by $3me/4$.  These explain the ambiguity in Eqs.~(\ref{HCNC1}) and (\ref{CC1}).

These results for the type I reproduce the corner charge formula for a cube-shaped crystal derived in Ref.~[\onlinecite{PhysRevB.102.165120}]. Furthermore, the fractional corner charge $\frac{\pm{e}}{8}$ in sodium chloride found in Ref.~[\onlinecite{watanabe2020fractional}] can be supported by our results as follows. Sodium chloride has a charge $\pm{e}$ at Wyckoff positions $1a$ and $3c$ and the opposite charge at Wyckoff positions $1b$ and $3d$ in the primitive unit cell. Thus, Eqs.~(\ref{BCNC}), (\ref{SCNC2}) and (\ref{HCNC1}) are satisfied and we determine the corner charge to be $\frac{\pm{e}}{8}$ from Eq.~(\ref{CC1}). 

\subsection{type I\hspace{-1pt}I: regular octahedron}\label{sectype2}
Here we consider a crystal in the shape of a regular octahedron with $n$ unit cells along each hinge, as shown in Fig.~{\ref{unitcell} (b)}.  
We can easily calculate the filling anomaly for one of the perfect crystals of type I\hspace{-1pt}I as shown in Fig.~\ref{perfectfig1} (b): 
\begin{eqnarray}
\eta_{n}^{\text{perfect,\ type\ I\hspace{-1pt}I}}&=&\frac{4}{3}n^{3}(\Delta a+\Delta b+3\Delta c+3\Delta d) \nonumber \\
&&+2n^{2}(\Delta a+3\Delta c) \nonumber \\ 
&&+\frac{2}{3}n(4\Delta a-2\Delta b+3\Delta c +3\Delta d) \nonumber \\
&&+\Delta a.
\end{eqnarray}
Again, the first and second terms can be dropped when the bulk and surfaces are charge neutral. Thus, from Eq.~(\ref{hinge_1_2}), we find
\begin{align}
\lambda_{\text{hinge}}^{\text{type\ I\hspace{-1pt}I}}&=\frac{4\Delta a-2\Delta b+3\Delta c+3\Delta d}{18}\times (-e)\notag\\
&=\frac{\Delta a+\Delta d}{2}e\ \left(\text{mod}\ \frac{e}{3}\right),
\label{HC2}
\end{align}
where we used Eqs.~(\ref{BCNC}) and (\ref{SCNC2}).  The charge neutrality condition for the hinge is thus
\begin{equation}
\label{HCNC2}
\Delta a +\Delta d\ \equiv 0\ (\text{mod}\ 2).
\end{equation}
Assuming Eq.~(\ref{HCNC2}) additionally and using Eq.~(\ref{corner_general}), we find the corner charge for type I\hspace{-1pt}I:
\begin{equation}
Q_{\text{corner}}^{\text{type\ I\hspace{-1pt}I}}=-\frac{\Delta a}{6}e\ \left(\text{mod}\ \frac{e}{3}\right).
\label{CC2}
\end{equation}

The ambiguity of the hinge charge  in Eq.~(\ref{HC2}) and the corner charge  in Eq.~(\ref{CC2}) for type I\hspace{-1pt}I crystals can be understood in the same way as in type I.
By attaching $C_{3}$-symmetric triangles with polarization charge $me/3$ (mod $e$) per period along the hinge on each surface, the hinge charge density is changed by an integer multiple of $e/3$ without affecting the bulk. Similarly, when the attached triangles have the corner charge $me/3$ (mod $e$), the corner charge of the octahedron is changed by $\frac{4}{3}me$.

Finally, there is an important point to note. While our results so far are derived when the center of the crystal is at Wyckoff position $1a$, they remain valid even when the center is at Wyckoff position $1b$ instead. This is because the change of the center from $1a$ to $1b$ is equivalent to exchanging Wyckoff positions $1a$ and $1b$, and $3c$ and $3d$. Our formulas in Eqs.~(\ref{HC2}) and (\ref{CC2}) are invariant under these exchanges, as long as the charge neutrality conditions are fulfilled.

\subsection{Ambiguities in hinge charges and corner charges}
\label{ApA}

\begin{figure}[t]
\centerline{\includegraphics[width=9cm,clip]{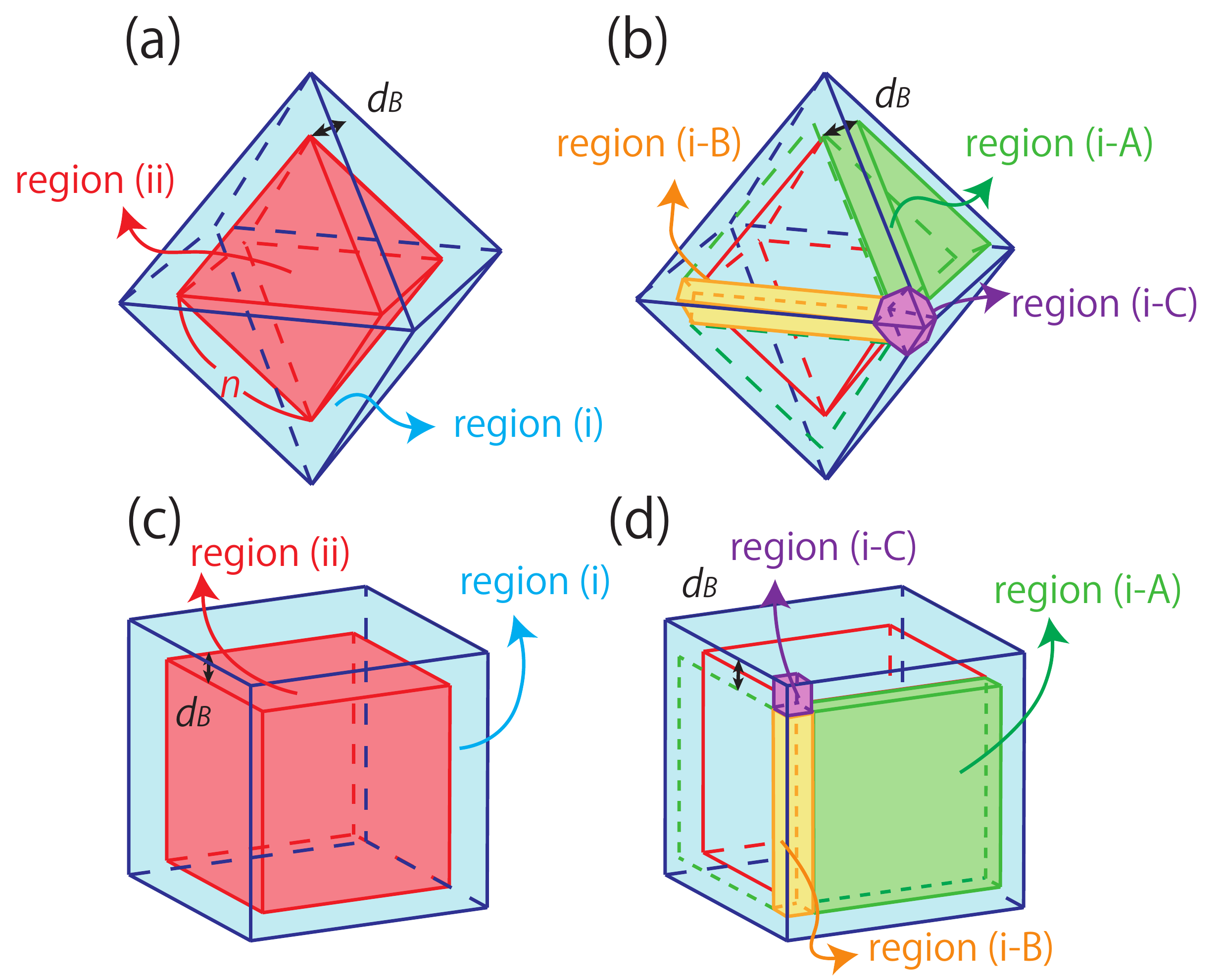}}
\caption{Conceptual pictures of the region division for (a,b) type I\hspace{-1pt}I and (c,d) type I. (a) The region (i\hspace{-1pt}i) is the inner octahedron. The region (i) is the remaining region. $d_{B}$ is taken to be large enough to regard region (i\hspace{-1pt}i) as a perfect crystal with $n$ periodicity. (b) A green regular triangular prism represents the region (i-A). The yellow region represents the region (i-B). The purple region represents the region (i-C). (c,d) For the cube, each region is defined in the same way as in (a) and (b).}
\label{ambiguityfig}
\end{figure} 

In this subsection, we identify the ambiguities in the hinge charge density and the corner charge for the type I and type I\hspace{-1pt}I crystals due to possible relaxation of electronic states and ionic positions near the boundaries, while the bulk electronic states and ionic positions are fixed in order to prove the ``modulo" parts of Eqs.~(\ref{HC1}), (\ref{CC1}), (\ref{HC2}) and (\ref{CC2}). We have already outlined how to obtain these ambiguities at the ends of Secs.~\ref{sectype1} and \ref{sectype2}; in this section we give more details to support their validity. We assume that all localized electronic states and ions in the crystal preserve the $432(O)$ symmetry, and that the periodicities along surfaces and hinges reflect those of the bulk. 

First of all, we begin with an ambiguity of the hinge charge for the type I\hspace{-1pt}I. In Sec.~\ref{sectype2}, we consider a perfect crystal of type I\hspace{-1pt}I and we get $\lambda^{L_{2}}=\frac{\Delta c+\Delta d}{2}e+\frac{\Delta b}{3}e$ under Eq.~(\ref{BCNC}). Meanwhile, if we consider type I\hspace{-1pt}I with possible relaxation, hinge charges may be modulated from those of the perfect crystal. The deviation of the hinge charge is generally expressed in multiples of some unit. We call this unit of the deviation of the hinge charges from those of the perfect crystal as an ambiguity of the hinge charge in the similar way to Ref.~[\onlinecite{PhysRevB.103.205123}].

From the definition of this ambiguity, we divide the total filling anomaly, $Q_{\text{tot}}$ into two parts, one of which is from the region (i) and the other is from region (i\hspace{-1pt}i) as shown in Fig.~\ref{ambiguityfig} (a): $Q_{\text{tot}}=Q_{\text{i}}+Q_{\text{i\hspace{-1pt}i}}$. The junctures of the region (i) and the region (i\hspace{-1pt}i) are taken so that the regions (i) and (i\hspace{-1pt}i) can be roughly regarded as surface and bulk regions respectively, and the thickness of region (i), $d_{\text{B}}$ is taken to be large enough so that electronic states and ionic positions in the region (i\hspace{-1pt}i) are the same as those in the bulk. In this paper, we impose an additional condition not included in Ref.~[\onlinecite{PhysRevB.103.205123}] that $d_{B}$ is independent of the size-parameter $n$,  in order to identify the order of $n$ in calculating the filling anomaly.

Given the charge neutrality conditions for the bulk and surfaces, the term proportional to $n$ in $Q_{\text{tot}}$ is evidently $12\lambda^{\text{type\ I\hspace{-1pt}I}}n/(-e)$. We now evaluate $Q_{\text{tot}}(=Q_{\text{i}}+Q_{\text{i\hspace{-1pt}i}})$ for the regions (i) and (i\hspace{-1pt}i) separately. First, the term proportional to $n$ in $Q_{\text{i\hspace{-1pt}i}}$ is $(-\frac{\Delta c+\Delta d}{2}-\frac{\Delta b}{3})\cdot 12n$ as shown in Sec.~\ref{sectype2}, since we take the region (i\hspace{-1pt}i) to be regarded as a perfect crystal.
Next, in order to consider the filling anomaly $Q_{\text{i}}$, we further divide the region (i) into three new regions (i-A), (i-B) and (i-C). We note that electronic states and ionic positions in the region (i) can be different from those in the bulk.

First, from $432(O)$ symmetry, we can define eight equivalent regions along the surfaces called regions (i-A) in the shape of a regular triangular prism preserving the $C_{3}$ symmetry, twelve equivalent regions along the hinges called regions (i-B) and the remaining six equivalent regions around the corners called regions (i-C) as shown in Fig.~\ref{ambiguityfig} (b). Here, we assign localized electronic orbitals and ions to each region so that their numbers are integers and they preserve the $432(O)$ symmetry. Given the charge neutrality condition for the surfaces, the charge included in region (i-A) is maximally in the $n^{1}$ order. Here, the region (i-A) can be regarded as a $C_{3}$-symmetric two-dimensional system with thickness $d_{B}$. In general, it is known that the bulk polarization for two-dimensional Wannier representable insulators with $C_{3}$ symmetry\cite{PhysRevB.99.245151} is quantized to an integer multiple of $\frac{e}{3s}(\bm{a}_{1}+\bm{a}_{2})$ (mod $\frac{e}{s}\bm{a}_{1},$ $\frac{e}{s}\bm{a}_{2}$), where $s$ is the area of the unit cell, and $\bm{a}_{1}$ and $\bm{a}_{2}$ are primitive lattice vectors. Therefore, its edge charge density is also quantized to an integer multiple of $e/3$. We conclude that the filling anomaly proportional to $n$ included in the eight equivalent regions (i-A) is $\frac{M}{3}\cdot n\cdot 3\cdot 8=8Mn$ where $M$ is an integer.

Second, since the region (i-B) has $\mathcal{O}(n)$ periodicity along the hinge, and each unit cell along the hinge in the region (i-B) contains an integer number of charge, the filling anomaly proportional to $n$ within the regions (i-B) is $12M'n$ where $M'$ is an integer. Finally the region (i-C) has no charge to the $n^{1}$ order. Thus, to summarize, the term proportional to $n$ in $Q_{\text{i}}$ is $8Mn+12M'n$.
Therefore, we conclude that $\lambda^{\text{type\ I\hspace{-1pt}I}}/(-e)=(-\frac{\Delta c+\Delta d}{2}-\frac{\Delta b}{3})+\frac{2M}{3}+M'$. The first and second terms on the right hand side of this equation comes from the bulk while the third and fourth terms can be an arbitrary integer multiple of $1/3$. Thus, the hinge charge density can be determined modulo $e/3$ only from the bulk and then $\lambda^{\text{type\ I\hspace{-1pt}I}}=-(\Delta c+\Delta d)e/2-\Delta be/3\equiv (\Delta a+\Delta d)e/2\ (\text{mod}\ e/3)$ holds.

Finally, we consider the ambiguity of the corner charge for the type I\hspace{-1pt}I under charge neutrality conditions for the bulk, surfaces and hinges. The numbers of localized electronic orbitals and ions in region (i) are generally even integers by $432(O)$ symmetry. Therefore, we get $6Q_{\text{corner}}/(-e)=\eta_{n}^{\text{perfect}}+2M$, where $M$ is an integer. This means we can determine the corner charge modulo $2e/6=e/3$ only from the bulk information. Thus, the ambiguity of the corner charge for type I\hspace{-1pt}I  is modulo $e/3$.

We briefly explain the ambiguities for the type I with reference to Figs.~\ref{ambiguityfig} (c) and (d). The filling anomaly proportional to $n$ coming from the region (i-A) is $\frac{Mn}{2}\cdot4\cdot6=12Mn$ ($M$: integer) because the bulk polarization for two-dimensional Wannier representable insulators with $C_{4}$ symmetry is quantized to an integer multiple of $(e/(2a),e/(2a))$ (mod $\frac{e}{a}$). Furthermore, the filling anomaly proportional to $n$ coming from the region (i-B) is $M'n\cdot12=12M'n$ ($M'$: integer). Therefore, to summarize, the term proportional to $n$ in the region (i) is $12(M+M')n$ leading to $(M+M')e\in \mathbb{Z}e$ ambiguity for the hinge charge density. Thus, the ambiguity of the hinge charge density for type I is modulo $e$. The numbers of localized electronic orbitals and ions in the region (i) are generally even integer by $432(O)$ symmetry. Thus, the corner charge in type I can be determined modulo $\frac{2}{8}e=\frac{1}{4}e$ only from the bulk information.

\subsection{type I\hspace{-1pt}I\hspace{-1pt}I: truncated cube}\label{sectype3}

So far, we have discussed the hinge and corner charges including their ambiguities in type I and type I\hspace{-1pt}I. Based on these results, we can determine those for types I\hspace{-1pt}I\hspace{-1pt}I-V including their ambiguities. In this subsection, we consider a crystal in the shape of a truncated cube as shown in Fig.~{\ref{unitcell} (c)}. There are two types of surfaces marked $S_{1}$ and $S_{2}$ and two types of hinges marked $L_{1}$ and $L_{3}$. The surface charge densities of $S_{1}$ and $S_{2}$ are clearly the same as those of type I and type I\hspace{-1pt}I, respectively. Thus, the charge neutrality condition for the surfaces is still Eq.~(\ref{SCNC2}). The $L_{1}$-hinge is equivalent to that of type I as long as the system size is large enough. Thus, the charge neutrality condition for the $L_{1}$-hinge is still Eq.~(\ref{HCNC1}).

\begin{figure}[h]
\centerline{\includegraphics[width=6cm,clip]{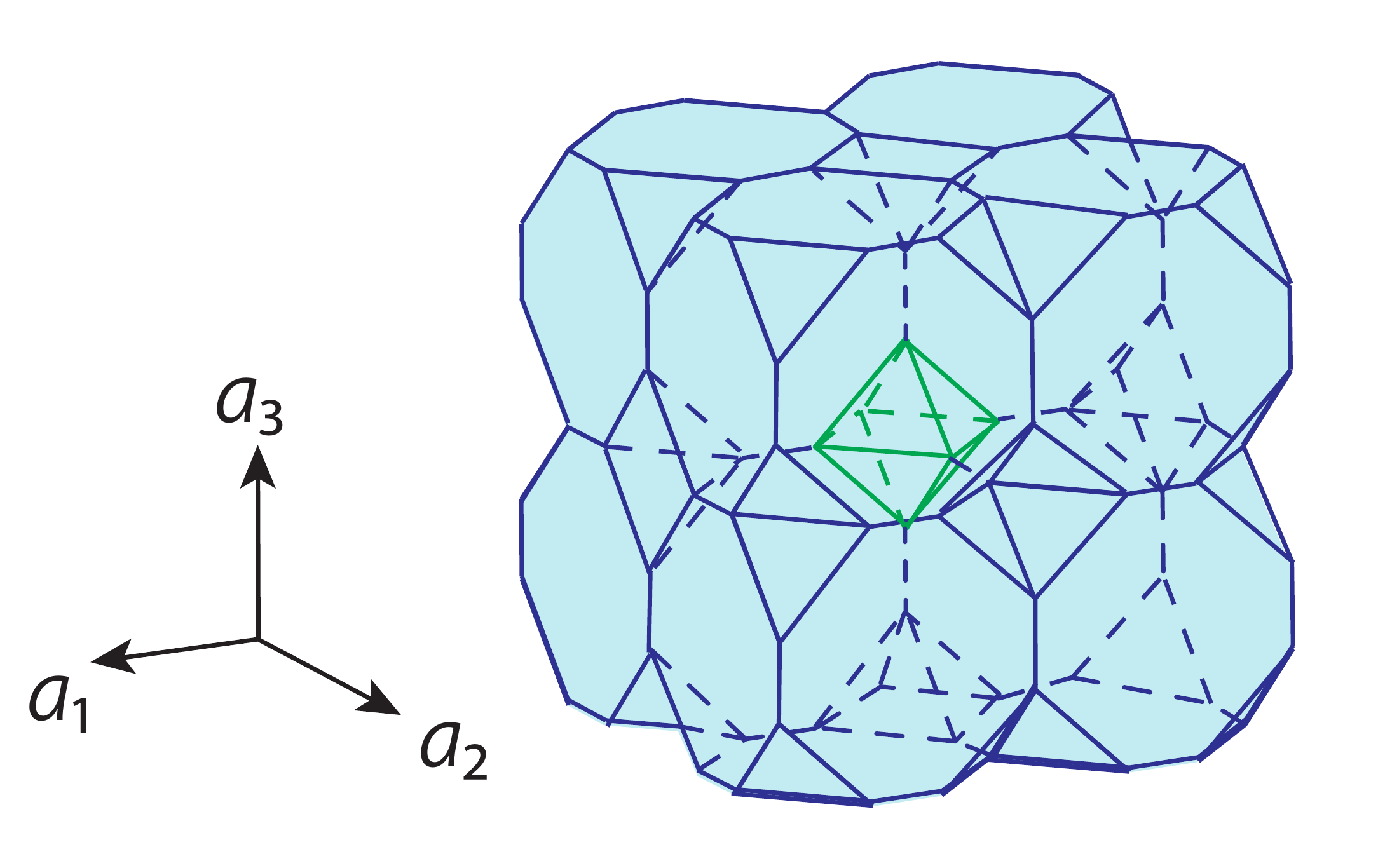}}
\caption{A schematic used for calculating the hinge and corner charges for type \mbox{I\hspace{-1pt}I\hspace{-1pt}I}. Eight equivalent type I\hspace{-1pt}I\hspace{-1pt}I crystals touch each other. The central green octahedron inside the entire system is a hollow space.}
\label{fusion}
\end{figure}

The remaining question is to derive the charge density for the $L_{3}$-hinge. To this end, we arrange eight crystals of \mbox{type I\hspace{-1pt}I\hspace{-1pt}I} which are exactly the same so that their surfaces touch each other as shown in Fig.~\ref{fusion}. Then, there is a hollow space with an octahedral shape of type I\hspace{-1pt}I at the center.
By filling this hollow space with a crystal of type \mbox{I\hspace{-1pt}I} and by locally modulating the electronic states and ionic positions near the boundaries of this type \mbox{I\hspace{-1pt}I} if needed, one can make the electronic states and ionic positions to be fully identical with those for the bulk of the original type \mbox{I\hspace{-1pt}I\hspace{-1pt}I}. We may need to add (or remove) electrons and ions at the interfaces, but this is already included in the ambiguities of the surface/hinge/corner charges. Since the bulk is charge neutral, the sum of the charges on the 24 equivalent $L_{3}$-hinges and on the 12 equivalent $L_{2}$-hinges is zero: $24\lambda^{L_{3}}+12\lambda^{L_{2}}=0$. Then, we get
\begin{equation}
\label{HCL3}
\lambda^{L_{3}}=-\frac{\Delta a+\Delta d}{4}e\equiv \frac{\Delta a+\Delta d}{4}e\ (\text{mod}\ \frac{e}{6}).
\end{equation}
Thus, the charge neutrality condition for the hinges in type I\hspace{-1pt}I\hspace{-1pt}I is $\Delta a+\Delta d\equiv0$ (mod 4) (same as Eq.~(\ref{HCNC1})). Similarly, the corner charge for type I\hspace{-1pt}I\hspace{-1pt}I is obtained by noting that the sum of the charges on the 24 equivalent corners in type I\hspace{-1pt}I\hspace{-1pt}I and on the 6 equivalent corners in type I\hspace{-1pt}I is zero: $24Q_{\text{corner}}^{\text{type I\hspace{-1pt}I\hspace{-1pt}I}}+6Q_{\text{corner}}^{\text{type I\hspace{-1pt}I}}=0$. Therefore, we get 
\begin{equation}
Q_{\text{corner}}^{\text{type I\hspace{-1pt}I\hspace{-1pt}I}}=\frac{\Delta a}{24}e\equiv -\frac{\Delta a}{24}e\ \left(\text{mod}\ \frac{e}{12}\right),
\end{equation}
under Eqs.~(\ref{BCNC}), (\ref{SCNC2}) and (\ref{HCNC1}).

\subsection{type I\hspace{-1pt}V: cuboctahedron}\label{sectype4}
In this subsection, we consider a crystal in the shape of a cuboctahedron as shown in Fig.~{\ref{unitcell} (d)}. Just as the discussion  in Sec.~\ref{sectype3}, the charge neutrality conditions for surfaces and hinges in type \mbox{I\hspace{-1pt}V} is Eq.~(\ref{SCNC2}) and Eq.~(\ref{HCNC1}), respectively. One can also easily calculate the corner charge from an observation that one corner of type \mbox{I\hspace{-1pt}V} can be regarded as a limit of merging two corners of type I\hspace{-1pt}I\hspace{-1pt}I together. Thus, we get
\begin{equation}
Q_{\text{corner}}^{\text{type I\hspace{-1pt}V}}=2Q_{\text{corner}}^{\text{type I\hspace{-1pt}I\hspace{-1pt}I}}=\frac{\Delta a}{12}e\equiv -\frac{\Delta a}{12}e\ \left(\text{mod}\ \frac{e}{6}\right),
\end{equation}
under Eqs.~(\ref{BCNC}), (\ref{SCNC2}) and (\ref{HCNC1}).

\subsection{type V: truncated octahedron}\label{sectype5}
In this subsection, we consider a crystal in the shape of a truncated octahedron as shown in Fig.~{\ref{unitcell} (e)}, by combining the results in Secs.~\ref{sectype2} and \ref{sectype3}.
Just as before, the charge neutrality condition for the surfaces in type V is still Eq.~(\ref{SCNC2}) and the charge neutrality condition for $L_{2}$-and $L_{3}$-hinges in type V is still Eq.~(\ref{HCNC1}). We note that the corner charge can be determined as $Q_{\text{corner}}^{\text{type V}}=-\frac{\eta_{n}}{24}e\ (\text{mod}\ \frac{e}{12})$ from Eq.~(\ref{corner_general}) under Eqs.~(\ref{BCNC}), (\ref{SCNC2}) and (\ref{HCNC1}), and the filling anomaly can be calculated only modulo 2 as a bulk quantity since the numbers of the electronic orbitals and ionic positions in the finite-sized crystal vary by any multiple of two except for those on the center of the crystal by possible relaxation presrving $432(O)$ symmetry. Therefore, we get $\eta_{n}=\Delta a$ (mod 2) and
\begin{equation}
Q_{\text{corner}}^{\text{type V}}=-\frac{\Delta a}{24}e\ \left(\text{mod}\ \frac{e}{12}\right),
\end{equation}
under Eqs.~(\ref{BCNC}), (\ref{SCNC2}) and (\ref{HCNC1}).

Finally, we summarize the real-space formulas for the hinge and corner charges in the five crystal shapes in Table~\ref{real-space-formulas}.
\onecolumngrid

\begin{table}[H]
\centering
\renewcommand{\arraystretch}{1.5}
\begin{tabular}{ccccccccc}
		\hline\hline
Type & $\lambda^{L_{1}}$ & $\lambda^{L_{2}}$ & $\lambda^{L_{3}}$ & $N_{\text{hinge}}^{L_{1}}$ & $N_{\text{hinge}}^{L_{2}}$ & $N_{\text{hinge}}^{L_{3}}$ & $Q_{\text{corner}}$ & $N_{\text{corner}}$ \\
\hline
I & $-\frac{\Delta a+\Delta d}{4}e$ (mod $e)$ & & & 12& & & $\frac{\Delta a}{8}e$ (mod $\frac{e}{4})$ & 8 \\

I\hspace{-1pt}I & & $\frac{\Delta a+\Delta d}{2}e$ (mod $\frac{e}{3})$ & & &$12$& & $\frac{\Delta a}{6}e$ (mod $\frac{e}{3})$ & 6 \\

I\hspace{-1pt}I\hspace{-1pt}I & $-\frac{\Delta a+\Delta d}{4}e$ (mod $e$) & & $\frac{\Delta a+\Delta d}{4}e$ (mod $\frac{e}{6})$ &$12$& &$24$& $\frac{\Delta a}{24}e$ (mod $\frac{e}{12}$) & 24   \\

I\hspace{-1pt}V & & & $\frac{\Delta a+\Delta d}{4}e$ (mod $\frac{e}{6})$ & & &$24$& $\frac{\Delta a}{12}e$ (mod $\frac{e}{6})$ & 12 \\

V & & $\frac{\Delta a+\Delta d}{2}e$ (mod $\frac{e}{3})$ & $\frac{\Delta a+\Delta d}{4}e$ (mod $\frac{e}{6})$ & & 12&24& $\frac{\Delta a}{24}e$ (mod $\frac{e}{12})$ & 24 \\
\hline \hline
\end{tabular}
\caption{The real-space formulas for the hinge and corner charges in the five crystal shapes. $N_{\text{hinge}}^{L_{i}}$ represents the number of the $L_{i}$-hinges in each type of the crystal ($i=1,2,3$). $N_{\text{corner}}$ represents the number of the corners in each type of the crystal.}
	\label{real-space-formulas}
\end{table}

\twocolumngrid

\section{formulation in terms of topological invariants}
\label{formulation by topological invariants}
In this section, we derive the hinge charge and corner charge formulas in terms of bulk band topology. Here, we first study spinless systems without time-reversal symmetry (TRS) in detail, and later systems with TRS and spinful systems.

 To this end, we use the method of the elementary band representation (EBR) matrix according to \mbox{Refs.~[\onlinecite{PhysRevB.103.165109}, \onlinecite{cano2021band}]}. The EBR matrix $A$ is an integer matrix with its columns representing an EBR and rows representing irreps of the $\bm{k}$ group of a high-symmetry point (HSP) in the BZ. 
Based on the above definition, we can connect the momentum-space representations of electronic states with those in real-space representations, since a group of topologically trivial bands can be expressed as a linear combination of EBRs. We can write
\begin{equation}
\label{EBRrelation}
v=A\tilde{n},
\end{equation}
where $v$ is the column vector with its $i$-th entry $v_{i}$ indicating the number of the $i$-th irrep of the $\bm{k}$ group of a HSP in the filled bands and $\tilde{n}$ is the column vector with its $i$-th entry $\tilde{n}_{i}$ indicating the number of the $i$-th irrep $\rho_{i}$ of the site-symmetry group of a Wyckoff position in the filled bands.

In order to obtain the number of Wannier orbitals localized at each Wyckoff position for a given band structure, we need to solve Eq.~(\ref{EBRrelation}) backwards; that is, our aim is to calculate $\tilde{n}$ from a given $v$ through Eq.~(\ref{EBRrelation}). 
The Smith normal form is useful to achieve this goal as shown in Refs.~[\onlinecite{PhysRevB.103.165109}, \onlinecite{cano2021band}]. The Smith normal form of $A$ is given by
\begin{equation}
\label{Smith}
A=U^{-1}DV^{-1},
\end{equation}
where $D$ is an integer matrix in the form of $D_{ij}=d_{i}\delta_{ij}$ ($d_{i}$; positive integer ($i=1,...,M$)), and $U$, $V$ are integer matrices invertible over integers. It is known that the most general solution of Eq.~(\ref{EBRrelation}) is 
\begin{equation}
\label{ntilde}
\tilde{n}=VD^{p}Uv+V\tilde{n}_{0},
\end{equation}
where $\tilde{n}_{0}$ is any vector in the null space of $D$, i.e., the first $M$ entries of $\tilde{n}_{0}$ are zero so that $D\tilde{n}_{0}=0$, and $D^{p}$ is the pseudoinverse matrix of $D$ which is made by transposing $D$ and then inverting the non-zero elements. Thus, given a particular $v$, $\tilde{n}_{i}$ can be determined only modulo gcd$\{ V_{ij}|_{j>M} \}$, where gcd indicates the greatest common divisor. when we need to calculate $n_{w}$, i.e., a sum of some $\tilde{n_{i}}$'s where $\rho_{i}$ is at the Wyckoff position $w$, we use
\begin{eqnarray}
\label{nw}
n_{w}&=&\sum_{i\in w}^{}\text{dim}(\rho_{i})(VD^{p}Uv)_{i} \nonumber \\
&& \left(\text{mod}\ \text{gcd}\left\{ \sum_{i\in w}^{}\text{dim}(\rho_{i})V_{ij}\right\}_{j>M} \right).
\end{eqnarray}
Furthermore, we can get $n_{a}+n_{b}+3n_{c}+3n_{d}$ in terms of the symmetry indicators with a modulo 
\begin{eqnarray}
\label{sum_nw}
&&\text{gcd}\left\{ \sum_{i\in a}^{}\text{dim}(\rho_{i})V_{ij}+\sum_{i\in b}^{}\text{dim}(\rho_{i})V_{ij}\right. \nonumber \\
&&\left.+3\sum_{i\in c}^{}\text{dim}(\rho_{i})V_{ij}+3\sum_{i\in d}^{}\text{dim}(\rho_{i})V_{ij}\right\}_{j>M} .
\end{eqnarray}
We can get $n_{b}+n_{d}$ and $n_{a}+n_{d}$ in terms of symmetry indicators in the same way. 

\begin{figure}
\centerline{\includegraphics[width=5cm,clip]{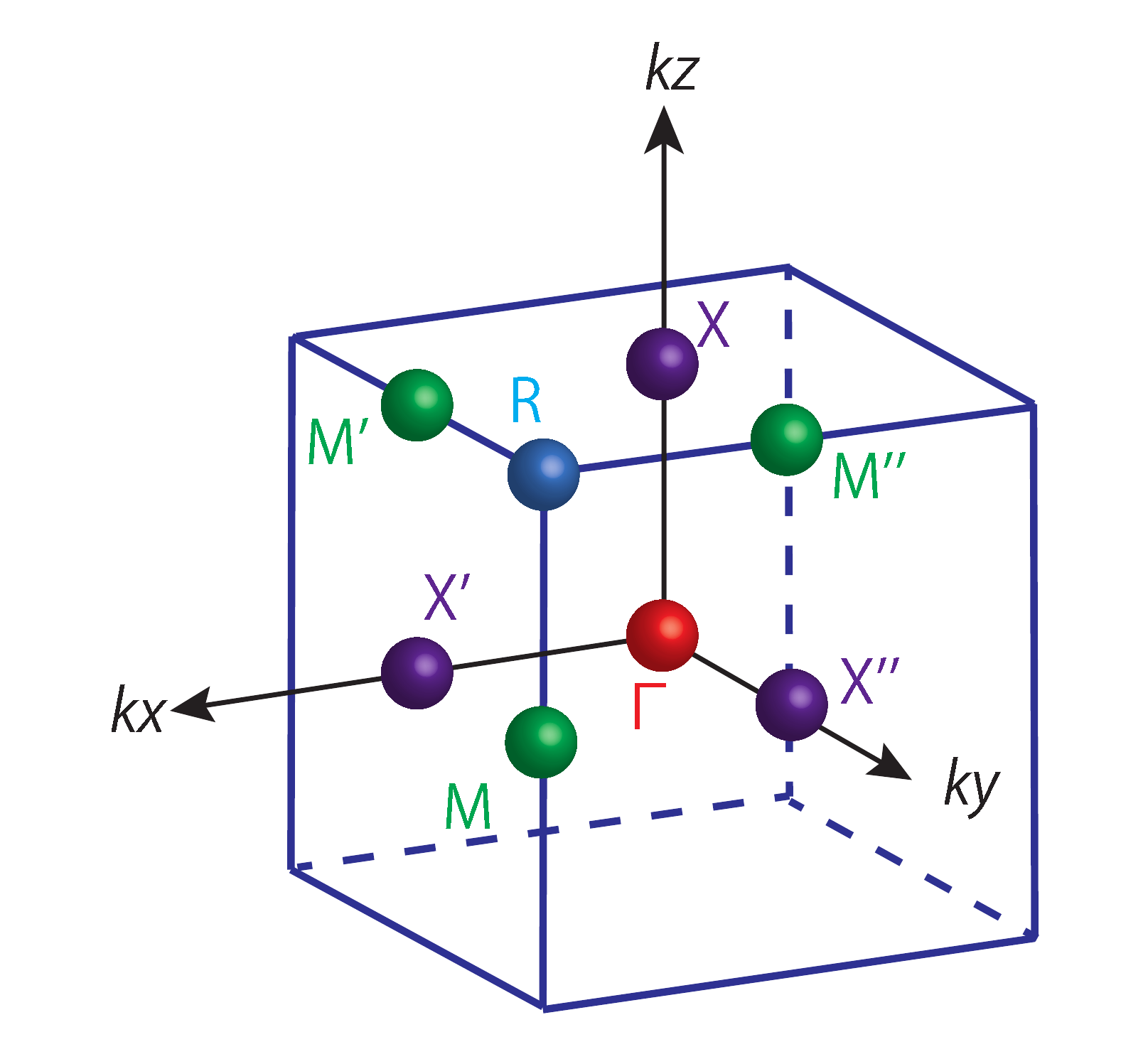}}
\caption{HSPs in momentum space with space group $P432$. The points with the same color belong to the same $\bm{k}$ vector star. $\Gamma=(0,0,0)$, $X=\frac{\bm{b}_{3}}{2}$, $M=\frac{\bm{b}_{1}+\bm{b}_{2}}{2}$ and $R=\frac{\bm{b}_{1}+\bm{b}_{2}+\bm{b}_{3}}{2}$, where $\bm{b}_{1}=(\pi /a,0,0)$, $\bm{b}_{2}=(0,\pi /a,0)$ and $\bm{b}_{3}=(0,0,\pi /a)$ are reciprocal lattice vectors. The blue cube represents the Brillouin Zone.}
\label{HSP.Fig}
\end{figure} 
In the following, we calculate the number of Wannier orbitals localized at each Wyckoff position (shown in Fig.~{\ref{unitcell} (a)}) from a given band representation with the space group $P432$. The site-symmetry groups of $1a$ and $1b$ are isomorphic to the point group 432($O$) and those of $3c$ and $3d$ are isomorphic to the point group 422($D_{4}$). In momentum space, there are four kinds of HSPs $\Gamma$, $R$, $M$, $X$ (shown in Fig.~{\ref{HSP.Fig}}) in the space group $P432$. The little groups at $\Gamma$ and $R$ are isomorphic to the point group $O$ and those at $M$ and $X$ are isomorphic to the point group $D_{4}$. The irreps of $O$ and $D_{4}$ are listed in Table~\ref{tab:432} and Table~\ref{tab:422}, respectively. Since the band representations induced from the other non-maximal Wyckoff positions such as $6e$, $6f$ and so on can be decomposed to direct sums of EBRs induced  from maximal Wyckoff positions, we need to consider only the band representations induced from maximal Wyckoff positions. 
\begin{table}[H]
\begin{center}
\begin{tabular}{cccccc}
		\hline\hline
432 ($O$) & $E$ & $C_{4z}$ & $C_{2x}$ & $C_{3\ (111)}$ & $C_{2\ (110)}$  \\
		\hline
$A_{1}$ & 1 & 1 & 1 & 1 & 1  \\
$A_{2}$ & 1 & $-1$ & 1 & 1 & $-1$  \\
$E$ & 2 & 0 & 2 & $-1$ & 0  \\
$T_{1}$ & 3 & 1 & $-1$ & 0 & $-1$  \\
$T_{2}$ & 3 & $-1$ & $-1$ & 0 & 1  \\
\hline \hline
\end{tabular}
\end{center}
\caption{Character table of the point group 432 ($O$).
}
	\label{tab:432}
\end{table}
\begin{table}[H]
\begin{center}
\begin{tabular}{cccccc}
		\hline\hline
422 ($D_{4}$) & $E$ & $C_{2z}$ & $C_{4z}$ & $C_{2x}$ & $C_{2\ (110)}$  \\
		\hline
$A_{1}$ & 1 & 1 & 1 & 1 & 1  \\
$A_{2}$ & 1 & $1$ & 1 & $-1$ & $-1$  \\
$B_{1}$ & 1 & 1 & $-1$ & $1$ & $-1$  \\
$B_{2}$ & 1 & 1 & $-1$ & $-1$ & $-1$  \\
$E$ & 2 & $-2$ & $0$ & 0 & 0  \\
\hline \hline
\end{tabular}
\end{center}
\caption{Character table of the point group 422 ($D_{4}$). Here, the $z$ axis is taken as the main fourfold rotational axis.
}
	\label{tab:422}
\end{table}
Each band representation is expressed as a vector $v$ in the basis:
\begin{eqnarray}
&&(A_{1}^{\Gamma},A_{2}^{\Gamma},E^{\Gamma},T_{1}^{\Gamma},T_{2}^{\Gamma},A_{1}^{R},A_{2}^{R},E^{R},T_{1}^{R},T_{2}^{R},A_{1}^{M}, \nonumber \\
&&A_{2}^{M},B_{1}^{M},B_{2}^{M},E^{M},A_{1}^{X},A_{2}^{X},B_{1}^{X},B_{2}^{X},E^{X})^{t},
\label{momentumspaceorder}
\end{eqnarray}
where $\rho^{\Pi}$ indicates the number of times the irrep $\rho$ appears in the given band representation at the HSP $\Pi$.
Each group of topologically trivial bands can be written as a linear combination of EBRs with integer coefficients. The coefficients form a vector $\tilde{n}$ in the basis:
\begin{eqnarray}
\label{AIs}
&&(A_{1}^{a},A_{2}^{a},E^{a},T_{1}^{a},T_{2}^{a},A_{1}^{b},A_{2}^{b},E^{b},T_{1}^{b},T_{2}^{b}, \nonumber \\
&&A_{1}^{c},A_{2}^{c},B_{1}^{c},B_{2}^{c},A_{1}^{d},A_{2}^{d},B_{1}^{d},B_{2}^{d})^{t},
\label{realspaceorder}
\end{eqnarray}
where $\rho^{w}$ indicates the number of times the EBR induced from the irrep $\rho$ of site-symmetry group of the Wyckoff position $w$ appears in the linear combination. It is important to note that $E^{c}$ and $E^{d}$ are not included in Eq.~(\ref{realspaceorder}), because both the band representations induced from $E^{c}$ and $E^{d}$ are composite, i.e., those can be continuously deformed into direct sums of  $T_{1}^{b}\oplus T_{2}^{b}$ and $T_{1}^{a}\oplus T_{2}^{a}$ via non-maximal Wyckoff positions $6f$ and $6e$, respectively.
In this basis, we construct the EBR matrix for the space group $P432$ according to the method of the band representation\cite{PhysRevB.97.035139, cano2021band, bradlyn2017topological,PhysRevX.7.041069} or by using Topological Quantum Chemistry in the Bilbao Crystallographic Server\cite{Aroyo:xo5013} as follows
\begin{equation}
\label{EBR432}
A = \left(
    \begin{array}{cccccccccccccccccc}
     1&0&0&0&0&1&0&0&0&0&1&0&0&0&1&0&0&0 \\
     0&1&0&0&0&0&1&0&0&0&0&0&1&0&0&0&1&0 \\
     0&0&1&0&0&0&0&1&0&0&1&0&1&0&1&0&1&0 \\
     0&0&0&1&0&0&0&0&1&0&0&1&0&0&0&1&0&0 \\
     0&0&0&0&1&0&0&0&0&1&0&0&0&1&0&0&0&1 \\
     1&0&0&0&0&0&1&0&0&0&0&0&0&1&0&1&0&0 \\
     0&1&0&0&0&1&0&0&0&0&0&1&0&0&0&0&0&1 \\
     0&0&1&0&0&0&0&1&0&0&0&1&0&1&0&1&0&1 \\
     0&0&0&1&0&0&0&0&0&1&0&0&1&0&1&0&0&0 \\
     0&0&0&0&1&0&0&0&1&0&1&0&0&0&0&0&1&0 \\
     1&0&1&0&0&0&0&0&0&1&0&0&0&1&1&1&0&1 \\
     0&1&1&0&0&0&0&0&1&0&0&1&0&0&0&1&1&1 \\
     0&0&0&1&0&0&1&1&0&0&0&1&1&1&0&1&0&0 \\
     0&0&0&0&1&1&0&1&0&0&1&1&0&1&0&0&0&1 \\
     0&0&0&1&1&0&0&0&1&1&1&0&1&0&1&0&1&0 \\
     1&0&1&0&0&0&0&0&1&0&1&0&0&0&1&1&1&0 \\
     0&1&1&0&0&0&0&0&0&1&0&0&1&0&1&0&1&1 \\
     0&0&0&1&0&1&0&1&0&0&1&1&1&0&1&0&0&0 \\
     0&0&0&0&1&0&1&1&0&0&1&0&1&1&0&0&1&0 \\
     0&0&0&1&1&0&0&0&1&1&0&1&0&1&0&1&0&1 \\
    \end{array}
  \right).
\end{equation}
Here, the $(i,j)$ component of $A$ indicates the number of times the $i$-th irrep of the momentum space appears in an EBR induced from the $j$-th irrep in the real space. For example, the $(1,1)$ component of $A$, i.e., ``1" means that the irrep $A_{1}$ of the little group of $\Gamma$ point appears once in the band representation induced from the irrep $A_{1}$ on the Wyckoff position $1a$.  

We can obtain the matrices $U$, $D$ and $V$ in Eq.~(\ref{Smith}) by applying Smith decomposition to this EBR matrix $A$ as follows
\onecolumngrid
\begin{equation}
\label{Umatrix}
U = \left(
    \begin{array}{cccccccccccccccccccc}
     1&0&0&0&1&0&0&0&0&0&0&0&0&-1&0&0&0&0&0&0 \\
     0&1&0&0&0&0&0&0&1&0&0&0&-1&0&0&0&0&0&0&0 \\
     0&0&0&0&0&0&0&1&0&0&0&0&0&0&0&0&0&0&0&0 \\
     0&0&0&0&0&0&0&0&1&0&0&0&0&0&0&0&0&0&0&0 \\
     0&0&0&0&1&0&0&0&0&0&0&0&0&0&0&0&0&0&0&0 \\
     1&0&0&1&0&0&0&1&-1&0&0&0&0&0&0&-1&0&0&0&0 \\
     0&0&0&0&0&0&0&0&-1&0&0&0&1&0&0&0&0&0&0&0 \\
     0&0&0&1&0&0&0&0&-1&0&0&0&0&0&0&0&0&0&0&0 \\
     -1&0&0&-1&-1&0&0&-1&1&0&0&0&0&1&0&1&0&0&0&0 \\
     1&0&0&0&0&0&0&1&0&1&0&0&0&-1&0&-1&0&0&0&0 \\
     -1&0&0&0&-1&0&0&-1&0&0&1&0&0&1&0&0&0&0&0&0 \\
     0&-1&0&-1&0&0&0&-1&0&0&0&1&1&0&0&0&0&0&0&0\\
     -1&-1&0&-1&0&0&1&-1&0&0&0&0&1&0&0&1&0&0&0&0 \\
     -1&0&0&0&-1&1&0&0&1&0&0&0&-1&1&0&0&0&0&0&0 \\
     1&0&0&0&0&0&0&1&-1&0&0&0&0&-1&1&-1&0&0&0&0 \\
     1&0&1&1&1&0&0&0&-1&0&0&0&0&-1&0&-1&0&0&0&0 \\
     0&-1&0&0&0&0&0&-1&-1&0&0&0&1&0&0&0&1&0&0&0 \\
     0&0&0&0&1&0&0&0&-1&0&0&0&0&-1&0&0&0&1&0&0 \\
     1&0&0&1&0&0&0&1&0&0&0&0&-1&-1&0&-1&0&0&1&0 \\
     0&0&0&-1&-1&0&0&0&0&0&0&0&0&0&0&0&0&0&0&1 \\
    \end{array}
  \right),
\end{equation}
\begin{equation}
\label{Dmatrix}
d_{1}=\cdots =d_{9}=1,
\end{equation}
\begin{equation}
\label{Vmatrix}
V = \left(
    \begin{array}{cccccccccccccccccc}
     1&0&0&0&0&0&0&0&0&-1&1&1&0&0&-1&0&0&0 \\
     0&1&0&0&0&0&0&0&0&-1&1&1&-1&1&-1&1&-1&0 \\
     0&0&1&0&0&0&0&0&0&0&-1&-1&0&-1&0&-1&0&-1 \\
     0&0&0&1&0&0&0&0&0&-1&0&0&-1&0&-1&0&0&0 \\
     0&0&0&0&1&0&0&0&0&-1&0&0&0&-1&0&0&0&-1 \\
     0&0&0&0&0&1&0&0&0&1&-1&-2&1&-1&1&-1&1&-1 \\
     0&0&0&0&0&0&1&0&0&1&-1&-1&0&-1&1&-1&0&0 \\
     0&0&0&0&0&0&0&0&0&0&1&0&0&0&0&0&0&0 \\
     0&0&0&0&0&0&0&1&0&1&0&-1&1&0&1&-1&0&0 \\
     0&0&0&0&0&0&0&0&0&1&0&0&0&0&0&0&0&0 \\
     0&0&0&0&0&0&0&0&1&0&0&1&-1&1&-1&1&-1&1 \\
     0&0&0&0&0&0&0&0&0&0&0&1&0&0&0&0&0&0 \\
     0&0&0&0&0&0&0&0&0&0&0&0&1&0&0&0&0&0 \\
     0&0&0&0&0&0&0&0&0&0&0&0&0&1&0&0&0&0 \\
     0&0&0&0&0&0&0&0&0&0&0&0&0&0&1&0&0&0 \\
     0&0&0&0&0&0&0&0&0&0&0&0&0&0&0&1&0&0 \\
     0&0&0&0&0&0&0&0&0&0&0&0&0&0&0&0&1&0 \\
     0&0&0&0&0&0&0&0&0&0&0&0&0&0&0&0&0&1 \\
    \end{array}
  \right).
\end{equation}
\twocolumngrid
We find from Eq.~(\ref{Vmatrix}) that the exact number of Wannier orbitals localized at each Wyckoff position, $n_{w}$, cannot be determined because the modulo part of Eq.~(\ref{nw}) becomes 1. Meanwhile, we can determine $n_{a}+n_{b}+3n_{c}+3n_{d}$ with no ambiguity which comes from Eq.~(\ref{Vmatrix}) by using Eqs.~(\ref{nw}) and (\ref{sum_nw}) as follows:
\begin{eqnarray}
\label{Birreps}
n_{a}+n_{b}+3n_{c}+3n_{d}&=&-A_{1}^{\Gamma}+A_{2}^{\Gamma}+T_{1}^{\Gamma}+T_{2}^{\Gamma}\nonumber \\ 
&&+2T_{1}^{R}+2B_{2}^{M}+2A_{1}^{X}.
\end{eqnarray}
In a similar way, we can get 
\begin{eqnarray}
\label{Sirreps}
n_{b}+n_{d}&=& A_{1}^{\Gamma}+4T_{1}^{\Gamma}+E^{R}-5T_{1}^{R}+B_{1}^{M}-A_{1}^{X}\nonumber \\
&=& A_{1}^{\Gamma}+E^{R}+T_{1}^{R}+B_{1}^{M}+A_{1}^{X} (\text{mod}\, 2),
\end{eqnarray}
\begin{eqnarray}
\label{Hirreps}
n_{a}+n_{d}&=& A_{1}^{\Gamma}+A_{2}^{\Gamma}+4T_{2}^{\Gamma}+2E^{R}\nonumber \\
&&+4T_{1}^{R}-B_{1}^{M}-B_{2}^{M} \nonumber \\
&=&A_{1}^{\Gamma}+A_{2}^{\Gamma}+2E^{R}\nonumber \\
&&-B_{1}^{M}-B_{2}^{M}\ (\text{mod}\, 4).
\end{eqnarray}
Equations~(\ref{Birreps}), (\ref{Sirreps}) and (\ref{Hirreps}) give the electronic contributions to the bulk, surface and hinge charges.
Therefore, the bulk, surface and hinge charge density can be obtained only from the symmetry indicators calculated from the occupied bands and ionic positions for the bulk under each ambiguity. In particular, the hinge charge formulas are
\begin{eqnarray}
\lambda^{L_{1}}&=&\frac{e}{4}\left\{m_{a}+m_{d}-\left(A_{1}^{\Gamma}+A_{2}^{\Gamma}+2E^{R}-B_{1}^{M}-B_{2}^{M}\right)\right\} \nonumber \\
&& \left(\text{mod}\, e\right),
\end{eqnarray}
\begin{eqnarray}
\lambda^{L_{2}}&=&\frac{e}{2}\left\{m_{a}+m_{d}-\left(A_{1}^{\Gamma}+A_{2}^{\Gamma}+2E^{R}-B_{1}^{M}-B_{2}^{M}\right)\right\} \nonumber \\
&& \left(\text{mod}\, \frac{e}{3}\right),
\end{eqnarray}
\begin{eqnarray}
\lambda^{L_{3}}&=&\frac{e}{4}\left\{m_{a}+m_{d}-\left(A_{1}^{\Gamma}+A_{2}^{\Gamma}+2E^{R}-B_{1}^{M}-B_{2}^{M}\right)\right\} \nonumber \\
&& \left(\text{mod}\, \frac{e}{6}\right).
\end{eqnarray}

On the other hand, we cannot determine the corner charge only from them, since we cannot determine $n_{w}$ modulo 2. Namely, each of the last four columns of the matrix $V$ shows that two systems with different Wyckoff positions of Wannier orbitals by an odd integer share the same irreps at all the HSPs. We note that such two systems cannot be continuously deformed into each other via non-maximal Wyckoff positions. We can distinguish those cases by using Wilson loops just as a similar approach is taken in two-dimensional systems\cite{PhysRevResearch.1.033074}. Here, we derive the momentum-space formulas for the corner charges by introducing the Wilson-loop invariants and then incorporating their values into the EBR matrix. In this paper, we define the Wilson loop\cite{PhysRevB.89.155114,PhysRevB.84.075119,PhysRevB.86.115112} as
\begin{equation}
W^{\gamma}=P\text{exp}\left(i\int_{\gamma}d\bm{k}\cdot \bm{A}(\bm{k})\right),
\end{equation} 
where $P$ represents the path-ordered product and we define the Berry connection\cite{PhysRevLett.62.2747} to be $[\bm{A}(\bm{k})]_{mn}=i\bra{u_{m}(\bm{k})}\nabla_{\bm{k}}\ket{u_{n}(\bm{k})}$ and $\ket{u_{n}(\bm{k})}$ to be a Bloch wave function with band index $n$ for occupied bands. The superscript $\gamma$ represents integral loops, $X$-$\Gamma$-$X$ line or $R$-$M$-$R$ line along the $k_{z}$ direction in the BZ. Here, the HSPs $\Gamma$, $X$, $M$ and $R$ represent $(0,0,0)$, $(0,0,\pi)$, $(\pi,\pi,0)$ and $(\pi,\pi,\pi)$ in the BZ, respectively.

Because $X$-$\Gamma$-$X$ line and $R$-$M$-$R$ line are invariant under $C_{4z}$ rotation, we can define the Wilson loop as a block diagonal form in terms of $C_{4z}$ eigenvalues: $W^{\gamma}=\bigoplus_{\alpha}W_{\alpha}^{\gamma}$, where $\alpha=1,-1,+i,-i$ is a $C_{4z}$ eigenvalue. Since the Wilson loops are unitary, their eigenvalues are in the form of $e^{i2\pi \theta_{\alpha \, ,j_{\alpha}}^{\gamma}}$, where a phase $\theta_{\alpha \, ,j_{\alpha}}^{\gamma}$ is real and $j_{\alpha}$ represents the $j_{\alpha}$-th eigenvalue of $W_{\alpha}^{\gamma}$.
Then, we define $\xi_{\alpha}^{\gamma}$ as a sum of phases of the Wison-loop eigenvalues for a particular $C_{4z}$ eigenvalue $\alpha$: 
\begin{equation}
\label{xigamma}
\xi_{\alpha}^{\gamma}=\frac{1}{2\pi i}\text{tr Log }W_{\alpha}^{\gamma}=\sum_{j_{\alpha}}^{}\theta_{\alpha,\, j_{\alpha} }^{\gamma},
\end{equation}
where Log means taking a principal value between $-\frac{1}{2}<\theta_{\alpha \, ,j}^{\gamma}\leq \frac{1}{2}$. The physical meaning of this value is the total sum of the $z$ components of the positions of Wannier centers of occupied Wannier orbitals with $C_{4z}=\alpha$\cite{PhysRevB.96.245115}. We can distinguish between any two topologically distinct band structures, originating from the last four columns of the $V$ matrix by using the following indicator: 
\begin{equation}
\label{Xidefinition}
\Xi=\left(\xi_{+i}^{\Gamma X}+\xi_{-i}^{\Gamma X}+\xi_{+i}^{MR}+\xi_{-i}^{MR}\right) \ (\text{mod}\, 2).
\end{equation}

This new indicator $\Xi$ has two important properties: (i) $\Xi$ is quantized to $0$ or $1$ modulo 2 (the proof of this property is shown in Appendix \ref{ApC}), (ii) we can eliminate the ambiguities of $n_{w}$ which come from the last four columns of the matrix $V$ by using $\Xi$ (see Appendix \ref{ApD}). In addition, the property (i) implies that $\Xi$ is constant through the continuous deformations via non-maximal Wyckoff positions between two systems with Wannier orbitals localized at different Wyckoff positions. For example, a system with electrons at the Wyckoff position $6e$ including an electron with $C_{4z}=+i$ at $(0,0,x)$ has $\xi_{+i}^{\Gamma X}=\xi_{+i}^{MR}=x$, $\xi_{-i}^{\Gamma X}=\xi_{-i}^{MR}=-x$ and $\Xi=0$. This system can be continuously deformed via $x\to0$ to that with six electrons at the Wyckoff position $1a$ with its representation as $T_{1}^{a}\oplus T_{2}^{a}$. Then, we get $\xi_{+i}^{\Gamma X}=\xi_{+i}^{MR}=\xi_{-i}^{\Gamma X}=\xi_{-i}^{MR}=0$ and $\Xi=0$. On the other hand, it can be also continuously deformed via $x\to \frac{1}{2}$ to that with two electrons per each Wyckoff position $3d$ with an irrep $E^{d}$. Then, we get $\xi_{+i}^{\Gamma X}=\xi_{+i}^{MR}=\xi_{-i}^{\Gamma X}=\xi_{-i}^{MR}=\frac{1}{2}$ and $\Xi=0$ (mod 2). Therefore, $\Xi$ remains constant through this continuous deformation via a non-maximal Wyckoff position.

Then we found that $n_{w}$ can be determined modulo 2 by incorporating the indicator $\Xi$ into the EBR matrix $A$. We define a pseudo-EBR matrix $\tilde{A}$ from the EBR matrix $A$, by adding one row representing values of $\Xi$ modulo 2 calculated for each atomic insulator in the form of Eq.~(\ref{AIs}). Then, we apply the Smith decomposition to this pseudo-EBR matrix $\tilde{A}$ and calculate $n_{w}$ just as before. As a result, we can determine each $n_{w}$ (mod 2) as expected (see Appendix \ref{ApF}). In particular, we get a simple result $n_{c}=\Xi$ (mod 2). Then, the corner charge formulas in terms of bulk band structures and bulk ionic positions are
\begin{eqnarray}
Q_{\text{corner}}^{\text{type I\hspace{-1pt}I}}&=&\frac{m_{c}-\Xi}{6}e\ \left(\text{mod}\, \frac{e}{3}\right),\label{CF1}\\ 
Q_{\text{corner}}^{\text{type I}}&=&\frac{m_{c}-\Xi}{8}e\ \left(\text{mod}\, \frac{e}{4}\right),\label{CF2}\\
Q_{\text{corner}}^{\text{type I\hspace{-1pt}V}}&=&\frac{m_{c}-\Xi}{12}e\ \left(\text{mod}\, \frac{e}{6}\right),\label{CF3}\\
Q_{\text{corner}}^{\text{type I\hspace{-1pt}I\hspace{-1pt}I}}&=&Q_{\text{corner}}^{\text{type V}}=\frac{m_{c}-\Xi}{24}e\ \left(\text{mod}\, \frac{e}{12}\right).\label{CF4}
\end{eqnarray}
We note that there are various equivalent ways to express the corner charges, because $\Delta a=\Delta b=\Delta c=\Delta d$ (mod 2) under the charge neutrality conditions for the bulk, surfaces and hinges in the types I-V.

From here, we consider spinless systems where TRS is imposed. According to Topological Quantum Chemistry in the Bilbao Crystallographic Server\cite{Aroyo:xo5013}, all the band representations in the space group $P432$ with TRS are the same as those without TRS. The only difference is that the decomposable band representations $E^{c}$ and $E^{d}$ become elementary band representations. By using the new EBR matrix made by incorporating new columns corresponding to $E^{c}$ and $E^{d}$ into Eq.~(\ref{EBR432}) and then applying the Smith decomposition to the new EBR matrix, we can obtain 
\begin{equation}
\label{Birreps2}
n_{a}+n_{b}+3n_{c}+3n_{d}=A_{1}^{\Gamma}+A_{2}^{\Gamma}+2E^{\Gamma}+3T_{1}^{\Gamma}+3T_{2}^{\Gamma},
\end{equation}
\begin{eqnarray}
\label{Sirreps2}
n_{b}+n_{d}&=& A_{1}^{\Gamma}-E^{\Gamma}+3T_{1}^{\Gamma}-A_{1}^{R}\nonumber \\
&&+E^{R}-5T_{1}^{R}+2B_{1}^{M}\nonumber \\
&=&  A_{1}^{\Gamma}+E^{\Gamma}+T_{1}^{\Gamma}+A_{1}^{R}\nonumber \\
&&+E^{R}+T_{1}^{R}\, (\text{mod}\, 2),
\end{eqnarray}
\begin{eqnarray}
\label{Hirreps2}
n_{a}+n_{d}&=& A_{2}^{\Gamma}+3T_{2}^{\Gamma}+A_{1}^{R}+2E^{R}\nonumber \\
&& +5T_{1}^{R}-2B_{1}^{M}\nonumber \\
&=& A_{2}^{\Gamma}-T_{2}^{\Gamma}+A_{1}^{R}+2E^{R}\nonumber \\
&& +T_{1}^{R}+2B_{1}^{M}\, (\text{mod}\, 4).
\end{eqnarray}
Furthermore, while $\xi_{+i}^{\gamma}=\xi_{-i}^{\gamma}$ follows due to the additional TRS, $\Xi$ is not constrained by it (see Appendix \ref{TRSconstraint}). We also find that the four pairs of the topologically distinct systems sharing the same irreps in the BZ are unchanged even if TRS is additionally imposed, and that there are no additional pairs like them in the new matrix $V$. Thus, we can determine $n_{w}'$s (mod 2) by incorporating the $\Xi$ defined by Eq.~(\ref{Xidefinition}) into the new EBR matrix. As before, we can get $n_{c}=\Xi$ (mod 2) which is the same result as that without TRS. Therefore, the corner charge formulas, Eqs.~(\ref{CF1})-(\ref{CF4}) are unchanged if TRS is added.

Next, we discuss spinful systems. The double group of $O$ has two two-dimensional irreps $\bar{E}_{1}$ and $\bar{E}_{2}$ and a four-dimensional irrep $\bar{F}$. The double group of $D_{4}$ has two two-dimensional irreps $\bar{E}_{1}$ and $\bar{E}_{2}$. As before, we can make the EBR matrix for spinful systems by using these irreps and then apply the Smith decomposition to it. Accordingly we get 
\begin{equation}
n_{a}+n_{b}+3n_{c}+3n_{d}=2\bar{E}_{1}^{\Gamma}+2\bar{E}_{2}^{\Gamma}+4\bar{F}^{\Gamma},\label{spinfulB}
\end{equation}
\begin{equation}
n_{b}+n_{d}=2(\bar{E}_{1}^{\Gamma}-\bar{E}_{1}^{R})=0\ (\text{mod}\, 2),\label{spinfulS}
\end{equation}
\begin{equation}
n_{a}+n_{d}=2\bar{E}_{1}^{R}+2\bar{E}_{2}^{\Gamma}+4\bar{F}^{\Gamma}=2\bar{E}_{1}^{R}+2\bar{E}_{2}^{\Gamma}\ (\text{mod}\, 4).\label{spinfulH}
\end{equation}
These results hold both with and without TRS.

Furthermore, in the spinful case, all the irreps of the double group of $O$ and $D_{4}$ have even dimensions regardless of the TRS. Therefore, all $n_{w}$ are even numbers and the corner charge is given by dividing $m_{a}$ by the number of corners. Namely, we get
\begin{equation}
Q_{\text{corner}}=\frac{m_{a}}{N_{\text{corner}}}e\, \left(\text{mod}\, \frac{2e}{N_{\text{corner}}}\right),\label{spinfulQc}
\end{equation}
for spinful systems both with and without TRS.
Thus, the corner charge has no electronic contribution for spinful systems.

Here we comment on different choices of definitions for the modulus for time-reversal symmetric systems with the spin-orbit coupling. This modulus, i.e., the ambiguity of the boundary charges, comes from a degree of freedom for the attached lower dimensional systems, and a set of the allowed lower dimensional systems is different between literatures. In Ref.~[\onlinecite{PhysRevResearch.1.033074}], the total number of electrons is doubled due to the Kramers degeneracy, and each ionic charge is even by including the spin-orbit coupling. Therefore, the modulus is doubled. In contrast, in the present paper, as already discussed in Ref.~[\onlinecite{PhysRevB.102.165120}], the total number of electrons is doubled but the ionic charge can be odd or even. Because the ionic charges of the materials can be an odd number in the unit of the elementary charge (i.e. the atomic number can be an odd number), we adopt the latter convention, and the modulus is not doubled even when the spin-orbit coupling is included.

\section{Conclusion}
\label{CaD}
In this paper, we derived the hinge charge and the corner charge formulas for the five crystal shapes of vertex-transitive polyhedra such as a cube, an octahedron and a cuboctahedron with cubic symmetry in terms of both bulk Wyckoff positions and bulk band structures. In their derivation, we showed that there are ambiguities depending on the finite-sized crystal shapes due to possible relaxation of electronic states and ionic positions near the boundaries for the same bulk electronic states and ionic positions. The hinge charges and the corner charges are determined as bulk quantities within these ambiguities. The strong dependence of boundary charge signatures and their ambiguities on the crystal shape is an interesting feature not found in two-dimensional systems.

We took the method of the EBR matrix to obtain the hinge charge and the corner charge formulas in terms of irreps at HSPs. While the hinge charge formulas can be constructed solely from the symmetry indicators, we find that the corner charges cannot even without time-reversal symmetry. This is because some band structures with different Wyckoff positions share the same irreps at all the HSPs in the BZ. To solve this problem, we proposed a Wilson-loop invariant $\Xi$. By incorporating this invariant $\Xi$, we constructed the corner charge formulas in terms of bulk band structures.

Finally, we briefly discuss the application of our results to real materials. To search for real materials with a nontrivial corner charge based on our formulas, we need to evaluate the Wilson-loop invariant, which cannot be incorporated into a high-throughput material search. Meanwhile, in ionic crystals one can easily see the Wyckoff positions of charges in real space, by which the corner charge can be evaluated via our formulas. As proposed in Ref.~[\onlinecite{watanabe2020fractional}], NaCl gives a nontrivial corner charge $\pm{e}/8$ in a crystal with the shape of a cube. It means that NaCl with other crystal shapes (types I\hspace{-1pt}I-V) also has nontrivial fractional corner charges. Nonetheless, the most stable crystal shape of NaCl is a cube, and it might be experimentally challenging to realize other crystal shapes with types I\hspace{-1pt}I-V. Furthermore, we expect that calcium fluorite CaF$_{2}$ has a fractional hinge charge density $e/2$ (mod $e$) along $L_{1}$-hinge. The crystal of CaF$_{2}$ has four Ca$^{2+}$ ions at Wyckoff positions $1b$ and $3d$, and eight F$^{-}$ ions at Wyckoff positions $8g$ in a unit cell. Thus, we get $\Delta a=8$ (or 0), $\Delta b=-2$ (or $6$), $\Delta c=0$ and $\Delta d=-2$. In this case, Eqs.~(\ref{BCNC}), (\ref{SCNC2}) and (\ref{HCNC2}) are satisfied, which means the bulk is charge neutral and the surfaces, $L_{2}$-hinges and $L_{3}$-hinges can be charge neutral under their respective ambiguities. Meanwhile, the \mbox{$L_{1}$-hinges} have fractional charge density $\lambda^{L_{1}}=e/2$ (mod $e$) from Eq.~(\ref{HC1}). Another example with the same property is barium titanate BaTiO$_{3}$ with the perovskite structure. 

Although the formulas we derived are for the five crystal shapes with cubic symmetry, we expect that basically the same derivation works for other vertex-transitive crystals, such as a tetrahedron. Extending the analysis to other crystal shapes and deriving formulas for them will certainly help identify more material canditates. We leave this important development to future work.

\begin{acknowledgments}
This work was supported by JSPS KAKENHI Grant Numbers
JP18H03678 and JP20H04633 and by Elements Strategy to Form Core Research (TIES) from MEXT Grant Number JP-MXP0112101001. The work of H.W. is supported by JSPS KAKENHI Grant No. JP20H01825 and by JST PRESTO Grant No. JPMJPR18LA.
\end{acknowledgments}

\appendix{}

\section{Proof of the quantization of $\Xi$}
\label{ApC}
In this appendix, we will show that $\Xi$ defined in Eq.~(\ref{Xidefinition}) is quantized to an integer. For this purpose, we first consider constraints for the Wilson loop due to the $432(O)$ symmetry.

First of all, we will show that we can define a Wilson loop $W_{\alpha}^{\gamma}$ for each $C_{4z}$ eigenvalue $\alpha$.
Here, the Wilson loop is described in terms of the projection operator onto the filled bands as follows: 
\begin{equation}
\label{projection}
[W^{\gamma}]_{mn}=\Braket{u_{k_{z}=\pi}^{m}|\prod_{k_{z}=-\pi}^{k_{z}=\pi}\mathcal{P}(k_{z})|u_{k_{z}=-\pi}^{n}},
\end{equation}
where the projection operator onto the filled bands $\mathcal{P}(k_{z})=\sum_{j\in \text{OCC}}^{}\ket{u_{k_{z}}^{j}}\bra{u_{k_{z}}^{j}}$ (OCC: occupied states) and $\prod_{k_{z}=-\pi}^{k_{z}=\pi}$ represents a path-ordered product from $k_{z}=-\pi$ to $k_{z}=\pi$ when an appropriate mesh is taken. We take the path $\gamma$ to be a $C_{4z}$-invariant line, either the $\Gamma X$ line ($k_{x}=k_{y}=0$) or the $MR$ line ($k_{x}=k_{y}=\pi$). Because the Bloch wave-number $\bm{k}$ is on the path $\gamma$, the Bloch Hamiltonian $h(\bm{k})$ and the fourfold rotational operator $C_{4z}$ around $z$ axis commute: $[h(\bm{k}), C_{4z}]=0$.
Therefore, we can express the Wilson loop in a block-diagonal form as follows: 
\begin{equation}
\label{diagonalWilsonloop}
W^{\gamma} = \left(
    \begin{array}{cccc}
     W_{1}^{\gamma}&&&\\
     &W_{-1}^{\gamma}&&\\
     &&W_{i}^{\gamma}&\\
     &&&W_{-i}^{\gamma}\\
    \end{array}
  \right),
\end{equation}
where we define $W_{\alpha}^{\gamma}$ as a Wilson loop within the $C_{4z}=\alpha$ sector.
Since the Wilson loop is unitary, $W_{\alpha}^{\gamma}$ is also unitary. 

Next, we will show that the sets of the phases $\theta_{\alpha, \, j_{\alpha}}^{\gamma}$ of the eigenvalues for $W_{\alpha}^{\gamma}$ satisfy the following relations due to the $432(O)$ symmetry: 
\begin{equation}
\label{constraintsC1}
\sum_{j_{\alpha}}^{}\theta_{\alpha, \, j_{\alpha}}^{\gamma}=\frac{M}{2}\ (\alpha=1,-1),
\end{equation}
\begin{equation}
\label{constraintsC2}
\sum_{j_{+i}}^{}\theta_{+i, \, j_{+i}}^{\gamma}+\sum_{j_{-i}}^{}\theta_{-i, \, j_{-i}}^{\gamma}=M',
\end{equation}
where $M$ and $M'$ are integers. Henceforth, we omit $\gamma$ if not specifically stated.

Because of $C_{2x}h(k_{z})C_{2x}^{-1}=h(-k_{z})$ from $C_{2x}$ symmetry, $C_{2x}\ket{u_{k_{z},\, \alpha_{n}}^{n}}$ is an eigenvector of $h(-k_{z})$, where $\ket{u_{k_{z},\, \alpha_{n}}^{n}}$ is the $n$-th eigenvector of $h(k_{z})$ with the $C_{4z}$ eigenvalue $\alpha_{n}$. Then, we can expand it by the Bloch eigenvectors of $h(-k_{z})$ as follows
\begin{equation}
\label{expand}
C_{2x}\ket{u_{k_{z},\, \alpha_{n}}^{n}}=\sum_{l\in \text{OCC}}^{}[B_{C_{2x}}(k_{z})]_{ln}\ket{u_{-k_{z},\, \alpha_{l}}^{l}}.
\end{equation}
By the orthonormality of the Bloch eigenstates, it is rewritten as
\begin{equation}
[B_{C_{2x}}(k_{z})]_{mn}=\Braket{u_{-k_{z},\, \alpha_{m}}^{m}|C_{2x}|u_{k_{z},\, \alpha_{n}}^{n}}.
\end{equation}
Here, the matrix $B_{C_{2x}}(k_{z})$ is unitary and is called a sewing matrix. 

From $C_{2x}C_{4z}C_{2x}^{-1}=C_{4z}^{-1}$, we derive
\begin{equation}
\label{ast}
C_{4z}(C_{2x}\ket{u_{k_{z},\, \alpha_{n}}^{n}})=\alpha_{n}^{\ast}(C_{2x}\ket{u_{k_{z},\, \alpha_{n}}^{n}}).
\end{equation}
By substituting Eq.~(\ref{expand}) to Eq.~(\ref{ast}), we derive
\begin{eqnarray}
\label{importantrelation}
\sum_{l\in \text{OCC}}^{}[B_{C_{2x}}(k_{z})]_{ln}(\alpha_{l}-\alpha_{n}^{\ast})\ket{u_{-k_{z},\, \alpha_{l}}^{l}}=0,
\end{eqnarray}
for any $n\in \text{OCC}$. By acting $\bra{u_{-k_{z},\, \alpha_{m}}^{m}}$ for any $m\in \text{OCC}$ to Eq.~(\ref{importantrelation}), we get $[B_{C_{2x}}(k_{z})]_{mn}(\alpha_{m}-\alpha_{n}^{\ast})=0$. Thus, if $[B_{C_{2x}}(k_{z})]_{mn}$ is nonzero, we get $\alpha_{m}=\alpha_{n}^{\ast}$. Therefore, $B_{C_{2x}}(k_{z})$ can be described in the same basis with Eq.~(\ref{diagonalWilsonloop}) as follows
\begin{equation}
\label{sewingmatrix}
B_{C_{2x}}(k_{z}) = \left(
    \begin{array}{cccc}
     B_{1,\, C_{2x}}&&&\\
     &B_{-1,\, C_{2x}}&&\\
     &&&B_{-i,\, C_{2x}}\\
     &&B_{+i,\, C_{2x}}&\\
    \end{array}
  \right),
\end{equation}
where $k_{z}$ dependence is omitted.
Since $B_{C_{2x}}(k_{z})$ is unitary, we can show that $B_{\alpha,\, C_{2x}}(k_{z})$ is also unitary by using Eq.~(\ref{sewingmatrix}). 

Given a symmetry of the system which transforms $\bm{k}$ into $\mathcal{O}_{\bm{k}}\bm{k}$, let $\mathcal{O}$ denote the matrix acting in the basis of the Bloch Hamiltonian. Then, the Bloch Hamiltonian $h(\bm{k})$ satisfies $\mathcal{O}h(\bm{k})\mathcal{O}^{-1}=h(\mathcal{O}_{\bm{k}}\bm{k})$. In this case, as is well known\cite{PhysRevB.96.245115}, the Wilson loop satisfies 
\begin{equation}
\label{sewingWilson}
B_{\mathcal{O}}(\bm{k})W^{\gamma}(\bm{k})B_{\mathcal{O}}^{\dag}(\bm{k})=W^{\mathcal{O}_{\bm{k}}\gamma}(\mathcal{O}_{\bm{k}}\bm{k}).
\end{equation}
By substituting Eqs.~(\ref{diagonalWilsonloop}) and (\ref{sewingmatrix}) to Eq.~(\ref{sewingWilson}), we can derive
\begin{equation}
\label{unitarytrans1}
B_{\alpha,\, C_{2x}}(k_{z})W_{\alpha}^{\gamma}B_{\alpha,\, C_{2x}}^{\dag}(k_{z})=W_{\alpha}^{-\gamma}=(W_{\alpha}^{\gamma})^{\dag},
\end{equation}
for $\alpha=1,-1$ and
\begin{equation}
\label{unitarytrans2}
B_{i,\, C_{2x}}(k_{z})W_{i}^{\gamma}B_{i,\, C_{2x}}^{\dag}(k_{z})=W_{-i}^{-\gamma}=(W_{-i}^{\gamma})^{\dag}.
\end{equation}
Thus, the set of eigenvalues of $W_{\alpha}^{\gamma}$ for $\alpha=1,-1$ is the same as its complex conjugation because $W_{\alpha}^{\gamma}$ and its own hermitian matrix are connected by a unitary transformation via Eq.~(\ref{unitarytrans1}). Therefore, we get
\begin{equation}
\label{complexconj1}
\{ \theta_{\alpha, \, j_{\alpha}}^{\gamma}\}=\{-\theta_{\alpha, \, j_{\alpha}}^{\gamma}\},
\end{equation}
for $\alpha=1,-1$ and for both of $\gamma$. Equation~(\ref{complexconj1}) says $\theta_{\alpha, \, j_{\alpha}}^{\gamma}$ is either constrained to 0, $\frac{1}{2}$ or forming a pair $\{+\theta_{\alpha, \, j_{\alpha}}^{\gamma},-\theta_{\alpha, \, j_{\alpha}}^{\gamma}\}$, so Eq.~(\ref{constraintsC1}) follows. In addition, the set of eigenvalues of $W_{i}^{\gamma}$ is the same as the complex conjugation of that of $W_{-i}^{\gamma}$ because $W_{i}^{\gamma}$ and $(W_{-i}^{\gamma})^{\dag}$ are connected by a unitary transformation via Eq.~(\ref{unitarytrans2}). Thus we get
\begin{equation}
\label{complexconj2}
\{ \theta_{+i, \, j}^{\gamma}\}=\{-\theta_{-i, \, j'}^{\gamma}\},
\end{equation}
for both of $\gamma$. Therefore, we can derive the Eq.~(\ref{constraintsC2}) from Eq.~(\ref{complexconj2}) if we note that the number of times $\frac{1}{2}$ appears in $\{ \theta_{+i, \, j}^{\gamma}\}$ is the same as that in $\{-\theta_{-i, \, j'}^{\gamma}\}$.

Finally, from the definition of $\Xi$ in Eq.~(\ref{Xidefinition}), which is
\begin{equation}
\label{Xi2}
\Xi=\sum_{j}^{}\left( \theta_{+i, \, j}^{\Gamma X}+\theta_{-i, \, j}^{\Gamma X}+\theta_{+i, \, j}^{MR}+\theta_{-i, \, j}^{MR} \right),
\end{equation}
it follows that $\Xi$ is quantized to an integer from Eq.~(\ref{constraintsC2}).

\section{Determination of the ambiguities of $n_{w}$ by using the value of $\Xi$}
\label{ApD}
In Sec.~\ref{formulation by topological invariants}, we have seen  that the corner charge cannot be fully determined only from the EBR matrix, because $n_{w}$ cannot be determined modulo 2. This ambiguity comes from the last four columns of the matrix $V$. 
In this appendix, we will show that one can remove this ambiguity by using the value of $\Xi$. Namely, $\Xi$ takes different values modulo 2 between two band structures which have the same irreps at every HSP in the BZ but different Wyckoff positions for the localized electronic states by an odd integer. These cases correspond to one of the four columns of the matrix $V$.
As an example of such bands, the seventeenth column of the matrix $V$ in Eq.~(\ref{Vmatrix}) means that the induced band representation from $A_{2}^{a}\oplus A_{1}^{c}$ and that from $A_{1}^{b}\oplus B_{1}^{d}$ have the same irreps at every HSP in the BZ. Nevertheless, they have distinct Wannier orbitals which cannot be continuously deformed into each other through non-maximal Wyckoff positions. Here, we show that $\Xi$ takes different values between these two cases.

First, we calculate the Wilson loop along both of the paths $\gamma$ for each $C_{4z}$ rotational eigenvalue to each atomic insulator represented by Eq.~(\ref{AIs}). Their phases represent Wannier centers along $z$ axis.
First of all, all the Wannier centers for the Wyckoff position $1a$ are zero, and thus $\Xi=0$.
The Wannier centers $\xi_{\alpha}^{\gamma}$ and $\Xi$ with respect to induced band representations from irreps of the site symmetry group of the other maximal Wyckoff positions $1b$, $3c$ and $3d$ are shown in Table~\ref{Wannier centers}. 
\begin{table}[H]
\begin{center}
\renewcommand{\arraystretch}{1.2}
\begin{tabular}{cccccccccc}
		\hline\hline
 irrep & $\xi_{+1}^{\Gamma X}$ & $\xi_{+1}^{MR}$ & $\xi_{-1}^{\Gamma X}$ & $\xi_{-1}^{MR}$ & $\xi_{+i}^{\Gamma X}$ & $\xi_{+i}^{MR}$ & $\xi_{-i}^{\Gamma X}$ & $\xi_{-i}^{MR}$ & $\Xi$ (mod 2) \\
		\hline
$A_{1}^{b}$ & $\frac{1}{2}$ & 0 & 0 & $\frac{1}{2}$ & 0 & 0 & 0 & 0 & 0 \\

$A_{2}^{b}$ & 0 & $\frac{1}{2}$ & $\frac{1}{2}$ & 0 & 0 & 0 & 0 & 0 & 0 \\

$E^{b}$ & $\frac{1}{2}$ & $\frac{1}{2}$ & $\frac{1}{2}$ & $\frac{1}{2}$ & 0 & 0 & 0 & 0 & 0 \\

$T_{1}^{b}$ & $\frac{1}{2}$ & 0 & 0 & $\frac{1}{2}$ & $\frac{1}{2}$ & $\frac{1}{2}$ & $\frac{1}{2}$ & $\frac{1}{2}$ & 0 \\

$T_{2}^{b}$ & 0 & $\frac{1}{2}$ & $\frac{1}{2}$ & 0 & $\frac{1}{2}$ & $\frac{1}{2}$ & $\frac{1}{2}$ & $\frac{1}{2}$ & 0 \\
\hline
$A_{1}^{c}$ & $\frac{1}{2}$ & 0  & $\frac{1}{2}$ & 0 & 0 & $\frac{1}{2}$ & 0 & $\frac{1}{2}$ & $1$ \\

$A_{2}^{c}$ & 0 & $\frac{1}{2}$ & 0 & $\frac{1}{2}$ & $\frac{1}{2}$ & 0 & $\frac{1}{2}$ & 0 & $1$ \\

$B_{1}^{c}$ & $\frac{1}{2}$ & 0 & $\frac{1}{2}$ & 0 & 0 & $\frac{1}{2}$ & 0 & $\frac{1}{2}$ & $1$ \\

$B_{2}^{c}$ & 0 & $\frac{1}{2}$ & 0 & $\frac{1}{2}$ & $\frac{1}{2}$ & 0 & $\frac{1}{2}$ & 0 & $1$ \\
\hline
$A_{1}^{d}$ & $\frac{1}{2}$ & $\frac{1}{2}$ & 0 & 0 & 0 & 0 & 0 & 0 &0 \\

$A_{2}^{d}$ & $\frac{1}{2}$ & $\frac{1}{2}$ & 0 & 0 & 0 & 0 & 0 & 0 &0 \\

$B_{1}^{d}$ & 0 & 0 & $\frac{1}{2}$ & $\frac{1}{2}$ & 0 & 0 & 0 & 0 &0 \\

$B_{2}^{d}$ & 0 & 0 & $\frac{1}{2}$ & $\frac{1}{2}$ & 0 & 0 & 0 & 0 &0 \\
\hline \hline
\end{tabular}
\end{center}
\caption{The values of $\xi_{\alpha}^{\gamma}$ and $\Xi$ for the irreps of the site symmetry group of  Wyckoff positions $1b$, $3c$ and $3d$.
}
	\label{Wannier centers}
\end{table}

Finally, we can show that $\Xi$ takes different values modulo 2 between two band structures related to the four columns of the matrix $V$ as shown in Table~\ref{detectationXi}, and thus we have removed the ambiguity by introducing $\Xi$.
\begin{table}[H]
\begin{center}
\renewcommand{\arraystretch}{1.2}
\begin{tabular}{ccc}
		\hline\hline
column number of $V$ & a pair of BRs & $\Xi$\\
\hline
\multirow{2}{*}{15} & $A_{1}^{a}\oplus A_{2}^{a}\oplus T_{1}^{a}\oplus A_{1}^{c}$ & $1$   \\
    & $A_{1}^{b}\oplus A_{2}^{b}\oplus T_{1}^{b}\oplus A_{1}^{d}$ & 0 \\
\hline
\multirow{2}{*}{16} & $E^{a}\oplus A_{1}^{b}\oplus A_{2}^{b}\oplus T_{1}^{b}$ & 0  \\
    & $A_{2}^{a}\oplus A_{1}^{c}\oplus A_{2}^{d}$ & $1$ \\
\hline
\multirow{2}{*}{17} & $A_{2}^{a}\oplus A_{1}^{c}$ & $1$   \\
    & $A_{1}^{b}\oplus B_{1}^{d}$ & 0 \\
\hline
\multirow{2}{*}{18} & $E^{a}\oplus T_{2}^{a}\oplus A_{1}^{b}$ & 0   \\
    & $A_{1}^{c}\oplus B_{2}^{d}$ & $1$ \\
\hline \hline
\end{tabular}
\end{center}
\caption{The values of $\Xi$ for two band structures related to the last four columns of the matrix $V$.
}
	\label{detectationXi}
\end{table}

\section{Smith normal form of the pseudo-EBR matrix $\tilde{A}$}
\label{ApF}
As we explained in Sec.~\ref{formulation by topological invariants}, we can introduce the pseudo-EBR matrix $\tilde{A}$ from the EBR matrix $A$ by adding a row $(0,0,0,0,0,0,0,0,0,0,1,1,1,1,0,0,0,0)$. This new row represents the values of $\Xi$ modulo 2 for each atomic insulator represented by Eq.~(\ref{AIs}) obtained from Table~\ref{Wannier centers}. The resulting pseudo-EBR matrix $\tilde{A}$ is
\onecolumngrid
\begin{equation}
\label{pseudoEBR}
\tilde{A} = \left(
    \begin{array}{cccccccccccccccccc}
     1&0&0&0&0&1&0&0&0&0&1&0&0&0&1&0&0&0 \\
     0&1&0&0&0&0&1&0&0&0&0&0&1&0&0&0&1&0 \\
     0&0&1&0&0&0&0&1&0&0&1&0&1&0&1&0&1&0 \\
     0&0&0&1&0&0&0&0&1&0&0&1&0&0&0&1&0&0 \\
     0&0&0&0&1&0&0&0&0&1&0&0&0&1&0&0&0&1 \\
     1&0&0&0&0&0&1&0&0&0&0&0&0&1&0&1&0&0 \\
     0&1&0&0&0&1&0&0&0&0&0&1&0&0&0&0&0&1 \\
     0&0&1&0&0&0&0&1&0&0&0&1&0&1&0&1&0&1 \\
     0&0&0&1&0&0&0&0&0&1&0&0&1&0&1&0&0&0 \\
     0&0&0&0&1&0&0&0&1&0&1&0&0&0&0&0&1&0 \\
     1&0&1&0&0&0&0&0&0&1&0&0&0&1&1&1&0&1 \\
     0&1&1&0&0&0&0&0&1&0&0&1&0&0&0&1&1&1 \\
     0&0&0&1&0&0&1&1&0&0&0&1&1&1&0&1&0&0 \\
     0&0&0&0&1&1&0&1&0&0&1&1&0&1&0&0&0&1 \\
     0&0&0&1&1&0&0&0&1&1&1&0&1&0&1&0&1&0 \\
     1&0&1&0&0&0&0&0&1&0&1&0&0&0&1&1&1&0 \\
     0&1&1&0&0&0&0&0&0&1&0&0&1&0&1&0&1&1 \\
     0&0&0&1&0&1&0&1&0&0&1&1&1&0&1&0&0&0 \\
     0&0&0&0&1&0&1&1&0&0&1&0&1&1&0&0&1&0 \\
     0&0&0&1&1&0&0&0&1&1&0&1&0&1&0&1&0&1 \\
     0&0&0&0&0&0&0&0&0&0&1&1&1&1&0&0&0&0\\
    \end{array}
  \right).
\end{equation}
We can obtain the matrices $\tilde{U}$, $\tilde{D}$ and $\tilde{V}$ in $\tilde{A}=\tilde{U}^{-1}\tilde{D}\tilde{V}^{-1}$ by applying the Smith decomposition to this pseudo-EBR matrix $\tilde{A}$ in the following. Accordingly, we can determine $n_{w}$ modulo 2 from symmetry indicators and the value of $\Xi$.

\begin{equation}
\label{tildeU}
\tilde{U} = \left(
    \begin{array}{ccccccccccccccccccccc}
     1&0&0&1&1&1&0&1&0&0&0&0&-1&-1&0&-1&0&0&0&0&1 \\
     1&1&0&1&1&0&0&1&0&0&0&0&-1&-1&0&-1&0&0&0&0&1 \\
     0&0&0&0&0&0&0&1&0&0&0&0&0&0&0&0&0&0&0&0&0 \\
     1&0&0&0&1&-1&0&0&0&0&0&0&1&-1&0&0&0&0&0&0&0 \\
     0&0&0&0&1&0&0&0&0&0&0&0&0&0&0&0&0&0&0&0&0 \\
     0&0&0&0&-1&0&0&0&0&0&0&0&0&1&0&0&0&0&0&0&-1 \\
     -1&0&0&0&-1&1&0&0&0&0&0&0&0&1&0&0&0&0&0&0&0 \\
     -1&0&0&1&-1&1&0&0&0&0&0&0&-1&1&0&0&0&0&0&0&0 \\
     0&0&0&0&0&0&0&0&0&0&0&0&0&0&0&0&0&0&0&0&1 \\
     0&0&0&1&0&1&0&1&0&0&0&0&-1&0&0&-1&0&0&0&0&1 \\
     -1&0&0&0&-1&0&0&-1&0&0&1&0&0&1&0&0&0&0&0&0&0 \\
     0&-1&0&-1&0&0&0&-1&0&0&0&1&1&0&0&0&0&0&0&0&0 \\
     -1&-1&0&-1&0&0&1&-1&0&0&0&0&1&0&0&1&0&0&0&0&0 \\
     1&0&0&0&0&0&0&1&0&1&0&0&0&-1&0&-1&0&0&0&0&0 \\
     0&0&0&0&-1&1&0&1&0&0&0&0&-1&0&1&-1&0&0&0&0&0 \\
     0&0&1&1&0&1&0&0&0&0&0&0&-1&0&0&-1&0&0&0&0&0 \\
     -1&-1&0&0&-1&1&0&-1&0&0&0&0&0&1&0&0&1&0&0&0&0 \\
     -1&0&0&0&0&1&0&0&0&0&0&0&-1&0&0&0&0&1&0&0&0 \\
     1&0&0&1&0&0&0&1&0&0&0&0&-1&-1&0&-1&0&0&1&0&0 \\
     0&0&0&-1&-1&0&0&0&0&0&0&0&0&0&0&0&0&0&0&1&0 \\
     -1&0&0&0&-1&1&0&0&1&0&0&0&-1&1&0&0&0&0&0&0&0 \\
    \end{array}
  \right),
\end{equation}
\begin{equation}
\label{tildeD}
\tilde{d}_{1}=\cdots=\tilde{d}_{10}=1,
\end{equation}
\begin{equation}
\label{tildeV}
\tilde{V} = \left(
    \begin{array}{cccccccccccccccccc}
     1&0&0&0&0&0&0&0&0&-1&1&-1&0&-1&-1&-1&0&-1 \\
     0&1&0&0&0&0&0&0&0&-1&1&-1&-1&0&-1&0&-1&-1 \\
     0&0&1&0&0&0&0&0&0&-1&-1&0&0&0&-1&-1&-1&-1 \\
     0&0&0&1&0&0&0&0&0&0&0&-1&-1&0&-1&0&0&0 \\
     0&0&0&0&1&0&0&0&0&-1&0&-1&0&-1&-1&-1&-1&-2 \\
     0&0&0&0&0&1&0&0&0&0&-1&1&1&1&0&0&0&0 \\
     0&0&0&0&0&0&1&0&0&0&-1&1&0&0&1&0&0&1 \\
     0&0&0&0&0&0&0&0&0&0&1&0&0&0&0&0&0&0 \\
     0&0&0&0&0&0&0&1&0&0&0&1&1&1&1&0&0&1 \\
     0&0&0&0&0&0&0&0&0&0&0&1&0&0&0&0&0&0 \\
     0&0&0&0&0&0&0&0&1&0&0&0&-1&0&0&1&0&1 \\
     0&0&0&0&0&0&0&0&0&0&0&0&0&-1&0&-1&0&-1 \\
     0&0&0&0&0&0&0&0&0&0&0&0&1&0&0&0&0&0 \\
     0&0&0&0&0&0&0&0&0&0&0&0&0&1&0&0&0&0 \\
     0&0&0&0&0&0&0&0&0&0&0&0&0&0&1&0&0&0 \\
     0&0&0&0&0&0&0&0&0&0&0&0&0&0&0&1&0&0 \\
     0&0&0&0&0&0&0&0&0&0&0&0&0&0&0&0&1&0 \\
     0&0&0&0&0&0&0&0&0&1&0&0&0&0&1&1&1&2 \\
    \end{array}
  \right).
\end{equation}
\twocolumngrid

\section{Proof of the relation of $\xi_{+i}^{\gamma}=\xi_{-i}^{\gamma}$ for systems with TRS}
\label{TRSconstraint}
We show that $\xi_{+i}^{\gamma}=\xi_{-i}^{\gamma}$ holds in spinless systems with TRS in this appendix. The derivation is similar to the one in Appendix \ref{ApC}.

First, we show a constraint for the Wilson loop due to TRS. Using the relation $\Theta h(k_{z})=h(-k_{z})\Theta$ from the TRS, where $\Theta$ is time-reversal operator, we can show that 
\begin{equation}
\label{expandV}
\Theta \ket{u_{k_{z},\, \alpha_{n}}^{n}}=\sum_{l\in \text{OCC}}^{}[V(k_{z})]_{ln}\ket{u_{-k_{z},\, \alpha_{l}}^{l}},
\end{equation}
where $\ket{u_{k_{z},\, \alpha_{n}}^{n}}$ is the $n$-th eigenvector of $h(k_{z})$ with the $C_{4z}$ eigenvalue $\alpha_{n}$.
By the orthonormality of the Bloch eigenstates, it is rewritten as
\begin{equation}
\label{sewingmatrixV}
[V(k_{z})]_{mn}=\Braket{u_{-k_{z},\, \alpha_{m}}^{m}|\Theta u_{k_{z},\, \alpha_{n}}^{n}}.
\end{equation}
Here, the matrix $V(k_{z})$ is unitary and is called a sewing matrix for TRS. We can show the following equation according to Ref.~[\onlinecite{PhysRevB.96.245115}]:
\begin{equation}
\label{sewingWilsonV}
V(\bm{k})\left(W^{\gamma}(\bm{k})\right)^{\ast}V^{\dag}(\bm{k})=W^{-\gamma}(-\bm{k})=\left(W^{\gamma}(\bm{k})\right)^{\dag}.
\end{equation}
The asterisk in the $\left(W^{\gamma}(\bm{k})\right)^{\ast}$ in Eq.~(\ref{sewingWilsonV}) denotes complex conjugation.
 
Second, we show a constraint for the sewing matrix $V(\bm{k})$ in terms of $C_{4z}$ eigenvalues. From Eq.~(\ref{expandV}), we get
\begin{equation}
\label{C4zV}
C_{4z}\Theta \ket{u_{k_{z},\, \alpha_{n}}^{n}}=\sum_{l\in \text{OCC}}^{}\alpha_{l}[V(k_{z})]_{ln}\ket{u_{-k_{z},\, \alpha_{l}}^{l}},
\end{equation}
where $\alpha_{l}$ is the $C_{4z}$ eigenvalue of the Bloch eigenvector corresponding to the $l$-th band.
From $C_{4z}\Theta=\Theta C_{4z}$, we can derive the following relation:
\begin{eqnarray}
\label{VC4z}
C_{4z}\Theta \ket{u_{k_{z},\, \alpha_{n}}^{n}}&=&\Theta C_{4z}\ket{u_{k_{z},\, \alpha_{n}}^{n}}\nonumber \\
&=&\Theta \alpha_{n}\ket{u_{k_{z},\, \alpha_{n}}^{n}} \nonumber \\
&=&\alpha_{n}^{\ast}\sum_{l\in \text{OCC}}^{}[V(k_{z})]_{ln}\ket{u_{-k_{z},\, \alpha_{l}}^{l}}.
\end{eqnarray}
By comparing Eq.~(\ref{C4zV}) with Eq.~(\ref{VC4z}), we get
\begin{eqnarray}
\label{importantrelation2}
\sum_{l\in \text{OCC}}^{}[V(k_{z})]_{ln}(\alpha_{l}-\alpha_{n}^{\ast})\ket{u_{-k_{z},\, \alpha_{l}}^{l}}=0,
\end{eqnarray}
for any $n\in \text{OCC}$. By acting $\bra{u_{-k_{z},\, \alpha_{m}}^{m}}$ for any $m\in \text{OCC}$ to Eq.~(\ref{importantrelation2}), we can get $[V(k_{z})]_{mn}(\alpha_{m}-\alpha_{n}^{\ast})=0$. Thus, if $[V(k_{z})]_{mn}$ is nonzero, we get $\alpha_{m}=\alpha_{n}^{\ast}$. This constraint is similar to the one on $B_{C_{2x}}(k_{z})$. Then, $V(k_{z})$ can be described in the same basis with Eq.~(\ref{diagonalWilsonloop}) as follows
\begin{equation}
\label{sewingmatrixV2}
V(k_{z}) = \left(
    \begin{array}{cccc}
     V_{1}&&&\\
     &V_{-1}&&\\
     &&&V_{-i}\\
     &&V_{+i}&\\
    \end{array}
  \right),
\end{equation}
where $k_{z}$ dependence is omitted.
Since $V(k_{z})$ is unitary, $V_{\alpha}(k_{z})$ is also unitary.

By substituting Eqs.~(\ref{diagonalWilsonloop}) and (\ref{sewingmatrixV2}) to Eq.~(\ref{sewingWilsonV}), we can derive
\begin{equation}
\label{unitarytransV2}
V_{i}(k_{z})\left(W_{i}^{\gamma}\right)^{\ast}V_{i}^{\dag}(k_{z})=W_{-i}^{-\gamma}=(W_{-i}^{\gamma})^{\dag}.
\end{equation}
Thus, the set of eigenvalues of $\left(W_{i}^{\gamma}\right)^{\ast}$ is the same as the complex conjugation of that of $W_{-i}^{\gamma}$ because $\left(W_{i}^{\gamma}\right)^{\ast}$ and $(W_{-i}^{\gamma})^{\dag}$ are connected by a unitary transformation via Eq.~(\ref{unitarytransV2}). Then, we can derive $\{ -\theta_{+i, \, j}^{\gamma}\}=\{-\theta_{-i, \, j'}^{\gamma}\}$. Namely, we get
\begin{equation}
\label{complexconjV2}
\{ \theta_{+i, \, j}^{\gamma}\}=\{\theta_{-i, \, j'}^{\gamma}\},
\end{equation}
for both of $\gamma$. From Eq.~(\ref{complexconjV2}), we can also derive  
\begin{equation}
\xi_{+i}^{\gamma}=\xi_{-i}^{\gamma}.
\end{equation}


%

\end{document}